\documentclass[aapm,prb,onecolumn,groupedaddress,showkeys]{revtex4-1} 

\usepackage[colorlinks=true,linkcolor=blue,citecolor=blue]{hyperref}

\usepackage{graphicx}
\usepackage{url}
\usepackage[utf8]{inputenc}
\usepackage{url}
\usepackage{color}
\usepackage[dvipsnames]{xcolor}
\usepackage{amsmath}
\usepackage{bbm}
\usepackage{amsfonts}
\usepackage{graphicx}
\usepackage{subfigure}
\usepackage{amssymb}%
\usepackage[normalem]{ulem}
\usepackage[toc,page]{appendix}
\usepackage{multirow}
\newcommand{\vecf}{f}
\newcommand{\vecw}{w}
\newcommand{\vecx}{x}

\newcommand{\tOpt}{t}

\begin{document}

\title{A linear programming approach to inverse planning in Gamma Knife radiosurgery}

\author{J. Sjölund\footnote{Corresponding author. Email: \href{mailto:jens.sjolund@elekta.com}{jens.sjolund@elekta.com}
}, S. Riad, M. Hennix, H. Nordström}
\affiliation{Elekta Instrument AB, Kungstensgatan 18, Box 7593, SE-103 93 Stockholm, Sweden}

\date{\today}

\begin{abstract}	

\textbf{Purpose:} Leksell Gamma Knife\textsuperscript{\textregistered} is a stereotactic radiosurgery system that allows fine-grained control of the delivered dose distribution. We describe a new inverse planning approach that both resolves shortcomings of earlier approaches and unlocks new capabilities.

\textbf{Methods:} We fix the isocenter positions and perform sector-duration optimization using linear programming, and study the effect of beam-on time penalization on the trade-off between beam-on time and plan quality. We also describe two techniques that reduce the problem size and thus further reduce the solution time: dualization and representative subsampling.

\textbf{Results:} 
The beam-on time penalization reduces the beam-on time by a factor 2--3 compared with the na\"{i}ve alternative. Dualization and representative subsampling each leads to optimization time-savings by a factor 5--20. Overall, we find in a comparison with 75 clinical plans that we can always find plans with similar coverage and better selectivity and beam-on time. In 44 of these, we can even find a plan that also has better gradient index. On a standard GammaPlan workstation, the optimization times ranged from 2.3 to 26 s with a median time of 5.7 s.

\textbf{Conclusion:} We present a combination of techniques that enables sector-duration optimization in a clinically feasible time frame. 

\end{abstract}

\maketitle

\section{Introduction}
Stereotactic radiosurgery (SRS) is defined as the use of externally generated ionizing radiation to inactivate or eradicate defined targets, typically in the head or spine, without the need for a surgical incision \cite{Barnett2007}. Present-day neurosurgeons routinely use stereotactic radiosurgery for the management of a wide variety of brain disorders, including certain malignant and benign tumors \cite{Kondziolka1991,Kondziolka1999,Lunsford2005,Sheehan2005}, as well as cardiovascular and functional disorders within the brain \cite{Kondziolka2012, Lunsford1991, Maesawa2001, Regis2006}.

Leksell Gamma Knife\textsuperscript{\textregistered} (LGK) is a dedicated system for intracranial stereotactic radiosurgery. Its most recent incarnations, Perfexion\texttrademark and Icon\texttrademark, use 192 $^{60}$Co sources, each emitting gamma radiation. The radiation is collimated to create a focus where the radiation from every source converges. At the focus, both the radiation intensity and its gradient become very large. This makes it possible to deliver high radiation doses with minimal damage to surrounding healthy tissue. 

The Perfexion\texttrademark \,and Icon\texttrademark \,systems enable two ways of tailoring the radiation dose according to the shape and size of the target. First, the patient can be precisely moved (robotically) in relation to the focus, effectively placing the focus in different \emph{isocenters}. Second, the radiation sources are arranged in eight, independently controlled, sectors. 
Each sector can be in one of four different collimator states: the 4 mm, 8 mm or 16 mm or in the beam-off state. For each isocenter position and collimator configuration (i.e.~collimator size for each sector), the irradiation time can be specified. This composition is often referred to as a {\it shot}.

The large number of degrees of freedom allows sculpting of the dose distribution in unparalleled ways. At the same time, however, it is infeasible to explore them all by means of manual planning. Thus, an inverse planning method is required to make the full potential of  LGK clinically accessible. Inverse planning methods only require the user to specify what objectives to strive for, and then uses mathematical optimization to search for the best possible treatment plan according to these objectives.

Inverse planning was introduced in Leksell GammaPlan\textsuperscript{\textregistered} 10, and has since then become widely adopted. This inverse planner uses well-established metrics such as coverage, selectivity and gradient index at a pre-determined isodose level, together with a beam-on time (BOT) penalization. The optimization variables are the position of isocenters, collimator configurations and irradiation times (beam weights more precisely). Unfortunately, the optimization problem is inherently difficult (non-convex). The difficulty arises from two components: that the objectives use relative isodoses (instead of absolute doses) and that the positions of the shots can change. Besides, the direct use of relative isodoses makes it difficult to simultaneously manage multiple targets and to enforce criteria such as the maximum dose to an organs at risk (OAR).

Researchers have proposed optimization approaches that use absolute doses and shots that are allowed to move \cite{Ferris2000,Ferris2002,Ferris2003, Ghaffari2012mip, Lee2000, Vandewouw2016}. This typically results in a so-called mixed-integer problem, which remains non-convex. In practice, this implies that there is a compromise between computation time and the risk of ending up in a suboptimal solution. A consequence is that it is difficult to explore what trade-offs are achievable---especially in complicated cases with multiple conflicting objectives. 

A remedy to most of these problems is to formulate a \emph{convex} optimization problem. Convexity is a highly desirable property that allows optimizations problems to be solved reliably and efficiently\cite{Boyd2004}. 
One way to achieve convexity---which we also use---is to fix the isocenter positions and perform sector-duration optimization \cite{Ghobadi2012}. This approach is inspired by optimization formulations, in particular fluence map optimization, that are common in intensity-modulated radiation therapy (IMRT) \cite{Bortfeld1999,Romeijn2008}. In sector-duration optimization, the collimator configurations are not packaged into shots during the optimization, instead the irradiation times of every collimator in every sector are treated independently during the optimization. The irradiation times are converted into deliverable shots after a solution has been found. Despite its promises, methods using sector-duration optimization has, so far, lacked an efficient way of controlling BOTs \cite{Ghaffari2012,Ghaffari2017} and have been too computationally costly for widespread clinical use \cite{Ghobadi2012,Ghobadi2013,Vandewouw2016}. This is about to change.

We present the first sector-duration optimization that uses linear programming, which is possible thanks to a well-founded
BOT penalization. Linear programming has a long history in radiotherapy\cite{Ehrgott2008,Holder2005,Reemtsen2009,Rosen1991,Shepard1999}, dating back at least to 1968\cite{Bahr1968}. Many of the early approaches constrained the doses to predetermined intervals and used the remaining freedom to maximize an objective function such as the mean\cite{Rosen1991} or minimum\cite{Lodwick1999} dose in the target. But, in practice, this upfront specification of dose constraints often lead to infeasible problems. This motivated the use of elastic constraints\cite{Holder2003}, where one instead minimizes the mean or maximum of the (weighted) constraint violations. Our formulation belongs to this category, but with an additional term for BOT penalization.  

Linear programming has also been used in radiosurgery before, specifically in two-phase approaches to inverse planning both for Gamma Knife radiosurgery\cite{Wu2003} and for robotic radiosurgery \cite{Schweikard1994,Schweikard1998,Kilby2010}. There, the decision variables are the weights of (deliverable) beams/shots that are generated in the first phase and optimized in the second.  In Gamma Knife radiosurgery, the sector-duration optimization has 24 times the number of decision variables of the corresponding two-phase procedure. The equivalent of a sector-duration optimization for a system with a multileaf collimator would include a decision variable for the left- and right positions of every leaf, resulting in a roughly hundredfold increase of decision variables. In this work, we present novel contributions that reduce the problem size of the sector-duration optimization while preserving the flexibility it offers. 
In summary, our method
\begin{itemize}
\item Manages multiple targets; 
\item Explicitly handles OARs;
\item Reproducibly finds the optimum given a fixed set of isocenter locations;
\item Runs in well under a minute;
\item Efficiently optimizes for short BOTs;
\item Allows hard constraints;
\item Allows the exploration of achievable trade-offs.
\end{itemize}

\section{Materials and Methods}
To achieve convexity, we divide the planning into three distinct phases: isocenter placement, optimization and sequencing. The isocenters chosen in the first phase remain fixed throughout the rest of the planning. In the optimization phase, we formulate an optimization problem where competing objectives are combined as a weighted sum. By changing weights it is straightforward to explore achievable trade-offs. Possible objectives include dose to target, sparing of OARs and a new---highly efficient---BOT penalization. During the optimization, times for each sector and collimator are allowed to vary independently. In the sequencing phase, these times are converted into deliverable shots.

\subsection{Isocenter placement}\label{sec:isocenters}
The first phase of the proposed inverse planner is to choose isocenter positions. These remain fixed throughout the subsequent phases, defining the search space of the optimization. The main objective of this phase is thus to provide enough freedom to find a high quality plan; the only harm in including extra isocenter positions is that it will take longer time to solve the optimization problem.

Although algorithms for automatic isocenter placement\citep{Wagner2000, John2005, Ghobadi2012, Bourland2000, Doudareva2015,Ferris2003} certainly have a role to play, we do not consider them to be the main focus of this work. Consequently, to remove this source of variability, we will reuse the isocenter positions from the corresponding (manual) reference plan in all examples that follow. 

\subsection{Optimization}
As we will describe below, a major difference between the inverse planner we propose and previous approaches based on sector-duration optimization\cite{Ghaffari2012,Ghaffari2017,Ghobadi2012,Ghobadi2013,Vandewouw2016} is that it uses \emph{linear programming}. Such optimization problems are well-behaved and well-studied \cite{Bertsimas1997,Boyd2004,Nocedal2006}. 

We recognize that there are multiple, possibly conflicting, objectives that are desirable. However, our working assumption is that the exact priority among these should be provided by the user. Consequently, we foresee that a number of options will be evaluated in the course of optimizing the treatment plan. This indeed reflects how it works presently, but the intention is that the proposed inverse planner will elucidate the achievable trade-offs. Formally, suppose we have $m$ cost functions, $\vecf=(f_1 (\vecx),\ldots,f_m (\vecx))^t$, representing e.g.~target dose, dose to an OAR and a penalization of BOT. This can be conveniently rearranged into a single cost function by forming a weighted sum, i.e.~by scalar multiplication with the weights $\vecw=(w_1,\ldots,w_m )^t$, where each weight quantifies the importance of the corresponding part of the cost function. We will now describe how we have defined the different parts of the optimization problem so that, taken together, it can be expressed as a linear programming problem.

\subsubsection{Dose-based objectives}
Given a fixed set of isocenter positions, our optimization variables are the irradiation times $\tOpt_{isc}$ corresponding to every isocenter $i$, collimator state $c$ and sector $s$. The dose $D$ at a position $r$ is linear in terms of these irradiation times, 
\begin{equation}
D(r, t_{isc}) = \sum_{i=1}^{N_{iso}}\sum_{s=1}^{8}\sum_{c=1}^{3}\Phi_{isc}(r)\tOpt_{isc}.
\end{equation}
Often, we are interested in the doses $D_n$ given to a discrete set of voxels at positions $r_n$, where $n=1,\ldots, N$. Then, after appropriate rearrangements, we may evaluate the dose as a matrix-vector multiplication: $D_n=(\Phi t)_n$. Thus, the $(N\times 24N_\text{iso})$-matrix $\Phi$, which we refer to as the dose rate kernel, maps irradiation times to doses.


In regions of interest, each voxel can be assigned dose-based objectives that reflect the amount of underdosage or overdosage it receives. Typically, the regions of interest include all targets, a volume of healthy tissue surrounding each target \cite{Ferris2000,Ghobadi2012} and clinically relevant OARs. We express the dose-based objectives using one or several hinge functions,
\begin{equation}
\left(D_n-\hat{D}_n\right)_+ = \text{max}\left(D_n-\hat{D}_n, 0\right),\label{eq:hinge}
\end{equation}
where $D_n$ is the dose in the voxel in question and $\hat{D}_n$ is a reference dose. This function, or its square, is commonly used as a dose-based objective\cite{Ghobadi2012,Romeijn2003}. In principle, every voxel could have a unique dose-based objective, which makes it possible to, for instance, perform full-fledged dose painting by numbers \cite{Alber2003, Bentzen2005, Ling2000}. By introducing auxiliary variables, piecewise linear convex functions such as the hinge function above, can be recast as linear programming problems, cf. appendix \ref{appendix:explicit_primal}.

\subsubsection{Beam-on time penalization}
To control the treatment time, we use a highly efficient penalization function which we refer to as the idealized beam-on time (iBOT) \cite{Ghaffari2012, Ghaffari2017}. It is defined as
\begin{equation}
\Theta(t) = \sum_{i=1}^{N_\text{iso}}\underset{s}{\max}\sum_{c=1}^3 t_{isc},
\end{equation}
where $N_\text{iso}$ is the number of isocenters; $s$ and $c$, respectively, correspond to the 8 sectors and 3 collimators. We emphasize that iBOT is \emph{not} an $L_p$-norm, nor does it promote sparsity---in fact, it is quite the opposite: it encourages the total BOT of each sector to be equally long, which is advantageous since they can irradiate simultaneously.
By introducing auxiliary variables, the iBOT function can be expressed using linear programming \cite{Svensson2014}, cf. appendix \ref{appendix:explicit_primal}.

\subsubsection{Constraints}
In a linear programming problem it is possible to include {\it hard constraints}, i.e.~conditions that the solution must fulfill. Physics only dictates one such hard constraint, namely that all times must be non-negative, $t\geq0$, and therefore this is the only hard constraint that is strictly necessary---exempting the ``artificial'' constraints coming from the auxiliary variables. One can, optionally, include hard constraints on doses in various regions.

\subsubsection{Normal tissue sparing}
We promote normal tissue sparing by encompassing the target within two non-overlapping thick ``shells'' shaped according to the target surface. By penalizing high doses inside the inner and outer shells we can control selectivity and gradient index (cf. section \ref{sec:metrics}), respectively. These two shells may seem similar to the widely used concept of margins\cite{ICRU2010}, but unlike margins---which have well-defined clinical meanings---our shells are merely used to steer the optimization. We construct the inner and outer shells as geometric expansions, defined via the Euclidean distance transform, of the target. The inner shell is expanded from the target surface until its total volume is half that of the target volume. The outer shell is expanded from the outer surface of the inner shell until its volume is twice that of the target.

\subsubsection{Illustrative example---primal formulation}
We will base our exposition on a minimal, yet illustrative, example of the optimization problem when there is a single target and an OAR where we want the limit the maximum dose to at most $D_{\rm O}$. In this case, the cost function has four components that control the target dose, selectivity, 
gradient index and BOT, respectively. We denote the prescription dose to the target $D_{\rm T}$ and the dose thresholds for selectivity and 
gradient index by $D_\textrm{S}$ and $D_\textrm{G}$, respectively.  We thus arrive at the  optimization problem
\begin{equation}\label{eq:costfcn}
\begin{aligned}
& \underset{t}{\text{minimize}} 
& & \frac{w_{\rm T}}{D_{\rm T} N_{\rm T}}\sum_{i=1}^{N_{\rm T}} (D_{\rm T}-(\Phi_{\rm T} \tOpt)_i)_++ \frac{w_{\rm S}}{D_{\rm S} N_{\rm S}}\sum_{i=1}^{N_{\rm S}}((\Phi_{\rm S} \tOpt)_i-D_{\rm S})_+ \\
&&&+\frac{w_{\rm G}}{D_{\rm G} N_{\rm G}} \sum_{i=1}^{N_{\rm G}}((\Phi_{\rm G} \tOpt)_i-D_{\rm G})_++\frac{w_{\rm BOT}}{D_\textrm{T}/\varphi_\text{cal}}\Theta(t)\\
& \text{subject to}& & \Phi_{\rm O} \tOpt\leq D_{\rm O}\\
&&& \tOpt\geq 0,
\end{aligned}
\end{equation}
where $\varphi_\text{cal}$ is the calibration dose rate,  $N_\alpha$ is the number of voxels in the structures $\alpha\in \{\textrm{T,\,S,\,G,\,O}\}$ and the weights $w_\alpha$ govern the relative importance of each term. Typically, we let $D_{\rm S} = D_{\rm T}$ to promote selectivity and $D_{\rm G} = D_{\rm T}/2$  to reduce the gradient index. In appendix \ref{appendix:explicit_primal}, we give the explicit representation of the problem in Eq.~\eqref{eq:costfcn} as a linear programming problem on standard form.

The example given here can easily be extended in several ways. For instance, additional structures could be incorporated as additional terms in the cost function, and homogeneous (target) dose distributions could be promoted by including a hinge function that penalizes overdosage.

\subsubsection{Representative subsampling}
The introduction of one auxiliary variable for every voxel in a relevant structure vastly increases the size of the optimization problem. In realistic cases, there could be, say, $10^5$ variables. Na\"ively solving the optimization problem thus becomes time-consuming at best, but it could even be impossible due to memory limitations. Thus, it is essential to reduce the size of the problem.

As a first step, we propose to use an approximation we refer to as \emph{representative subsampling}. Representative subsampling is based on the observation that the dose tends to vary quite smoothly from one position to another. True enough, there are regions like the penumbra region of the Gamma Knife (and flattening filter-free accelerators), that may exhibit stronger variability than others, but for practical purposes it is redundant to take the dose in every voxel into account during the optimization. We exploit this realization by sampling only a representative fraction of the voxels in each structure that we use in the optimization. 

Formally, we may understand this from the observation that most of the criteria involving a structure $\Omega$ can be phrased as an integral
\begin{equation}
J = \int_\Omega f\left(r,D(r,t))\right)dr\,,\label{eq:integral}
\end{equation}
where $f$ is a function that assigns a cost to the dose, $D(r,t)=\Phi(r)t$, at position $r$. For instance, a dose-based hinge functions, as in equation \eqref{eq:hinge}, corresponds to $f(r,D(r,t))=\left(D(r,t)-\hat{D}(r)\right)_+$ and a hard constraint on the maximum dose corresponds to 
\begin{equation}
f(r,D(r,t)) = \begin{cases}
0&\text{if }D(r,t)\leq \hat{D}(r),\\
\infty&\text{otherwise}.
\end{cases}
\end{equation}
If we discretize the integral \eqref{eq:integral} uniformly (i.e. at the sampling points) we recover the conventional expression
\begin{equation}
J\approx \frac{V(\Omega)}{N}\sum_{n=1}^N f(r_n, \Phi(r_n)t),
\end{equation}
where $V(\cdot)$ is the volume operator. However, it is well-known that this scales poorly with dimension---the number of samples required to reach a given precision increases exponentially with the dimension. To mitigate this, researchers have proposed to subsample voxels on a regular grid\cite{Ghaffari2012mip} or to aggregate voxels into clusters\cite{Scherrer2005,Ungun2018}. Another option is to use a stochastic gradient descent algorithm that resamples voxels at each iteration\cite{Martin2007}, but stochastic gradient descent is often ineffective on constrained problems, since every iteration requires a projection onto the feasible set \cite{Wang2016}.

Randomness is powerful, though. Instead of the deterministic sampling schemes mentioned above, we build upon the well-established efficiency of Monte Carlo integration to approximate Eq. \eqref{eq:integral} by sampling positions uniformly at random in the structure. In addition, because we want to subsample also when using minimum and maximum dose constraints in the target or maximum dose constraints in organs at risk, we include a separate term corresponding to positions sampled at random on the tessellated surface of the volume. 

\subsubsection{Dualization}
The introduction of auxiliary variables in the primal formulation increases the number of variables---often severalfold. However, as can be appreciated from the explicit formulation in appendix \ref{appendix:explicit_primal}, the resulting matrix is highly structured; it could e.g.~be decomposed as the sum of a low-rank matrix and a sparse matrix. Exploiting this structure makes it possible to reduce computation times drastically. We have found that dualization reduces the computation time by a factor 5--20 depending on the features of the problem. Since strong duality holds for linear programming problems, the primal and dual problems are equivalent. 

In appendix \ref{appendix:explicit_dual} we revisit our earlier example and give the explicit formulas for the corresponding dual problem. Importantly, however, the constraints introduced when rewriting the hinge function using auxiliary variables become trivial thanks to the dualization. The resulting size reduction leads to dramatic performance improvements when solving the dual compared to the original, primal, problem. Moreover, the computational gain due to dualization is entirely complementary to that of representative subsampling. 

\subsection{Sequencing algorithm for shot composition}
For the Gamma Knife system to deliver the treatment composite shots are required, i.e.~collimator and sector configurations for each shot. The conversion from irradiation times for each isocenter position $\tOpt_{isc}$ to such shots is referred to as sequencing. The sequencing can be done in several different ways, and although we will not go into details here \citep{nordstrom2015sequencing}, it is worth noting that the result of the optimization will often lead to multiple shots in the same isocenter position.\\
\begin{figure}[ht]
\centering
\includegraphics[scale=0.6]{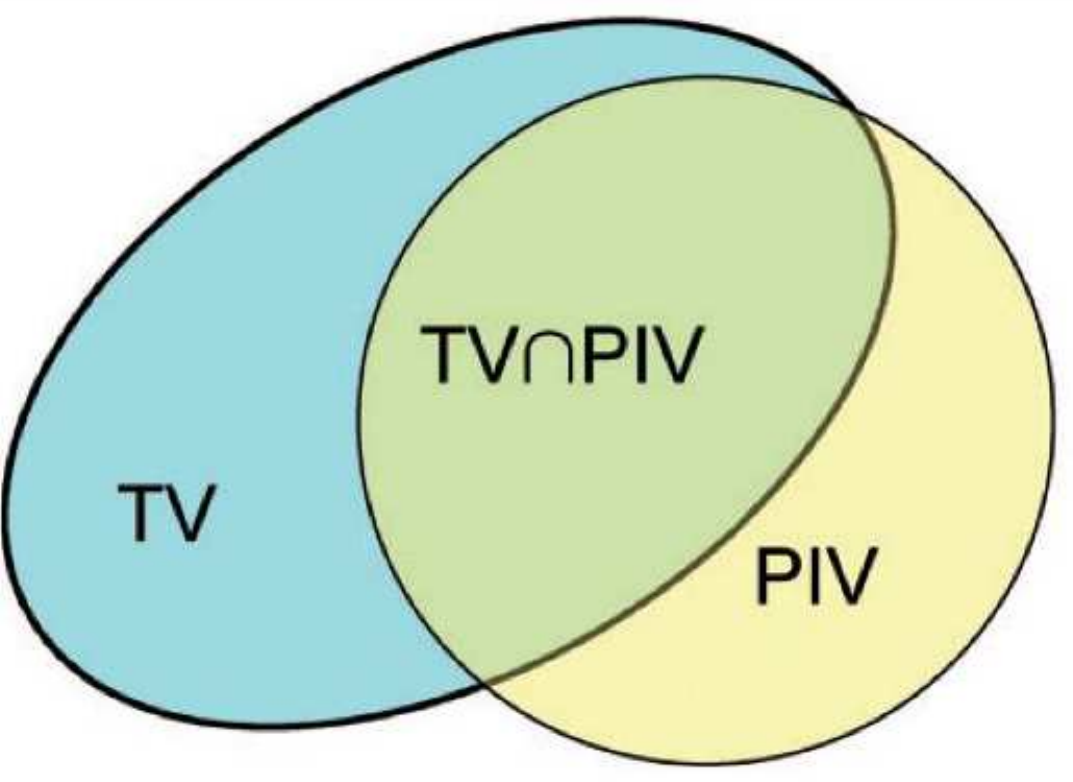}
\caption{The target volume (TV) and planning isodose volume (PIV)}\label{fig:TV}
\end{figure}
\subsection{Evaluation}\label{sec:metrics}
In Fig.~\ref{fig:TV}, we define the target volume (TV) and the planning isodose volume (PIV), which are used, together with the volume operator $V(\cdot)$, to define the common radiosurgery metrics\cite{Torrens2014}: 
\begin{eqnarray}
\textrm{Coverage:} & \hspace{4 mm} C = \frac{V(PIV \cap TV)}{V(TV)}\,, \\
\textrm{Selectivity:} & \hspace{4 mm} S = \frac{V(PIV \cap TV)}{V(PIV)}\,, \\
\textrm{Gradient index:} &\hspace{4 mm} GI = \frac{V(PIV_{ISO/2})}{V(PIV_{ISO})}\,.
\end{eqnarray}
In words, coverage is the fraction of the target volume that receives at least as high dose as prescribed, selectivity is the fraction of the planning isodose volume that encompasses the target volume, and gradient index describes the dose fall-off by the ratio of the volume that receives at least half the prescription dose to the planning isodose volume.
In addition to these metrics, plan quality is often evaluated based on the Paddick conformity index (PCI)\citep{Paddick2000}, which is defined as:
\begin{eqnarray}
\textrm{Paddick conformity index:} & \hspace{4 mm} PCI = C\cdot S.\label{eq:PI}
\end{eqnarray}

\section{Results}
We have evaluated both the overall performance of the proposed inverse planner as well as studied the novel components---the beam-on time penalization and the subsampling scheme---in isolation. We begin by reporting on the beam-on time penalization and the subsampling scheme. Then, we describe the overall performance on a set of 75 clinically acceptable reference plans. Finally, we consider one case in more detail. For this case, we illustrate the achievable trade-offs and show how to perform dose painting or create a homogeneous dose distribution. 

\subsection{Beam-on time penalization}\label{sec:BOT}
\begin{figure}%
 \centering
 \subfigure[Paddick index vs. BOT]{%
\label{fig:iBOT_vs_BOT1}%
 \includegraphics[width = 0.5\textwidth]{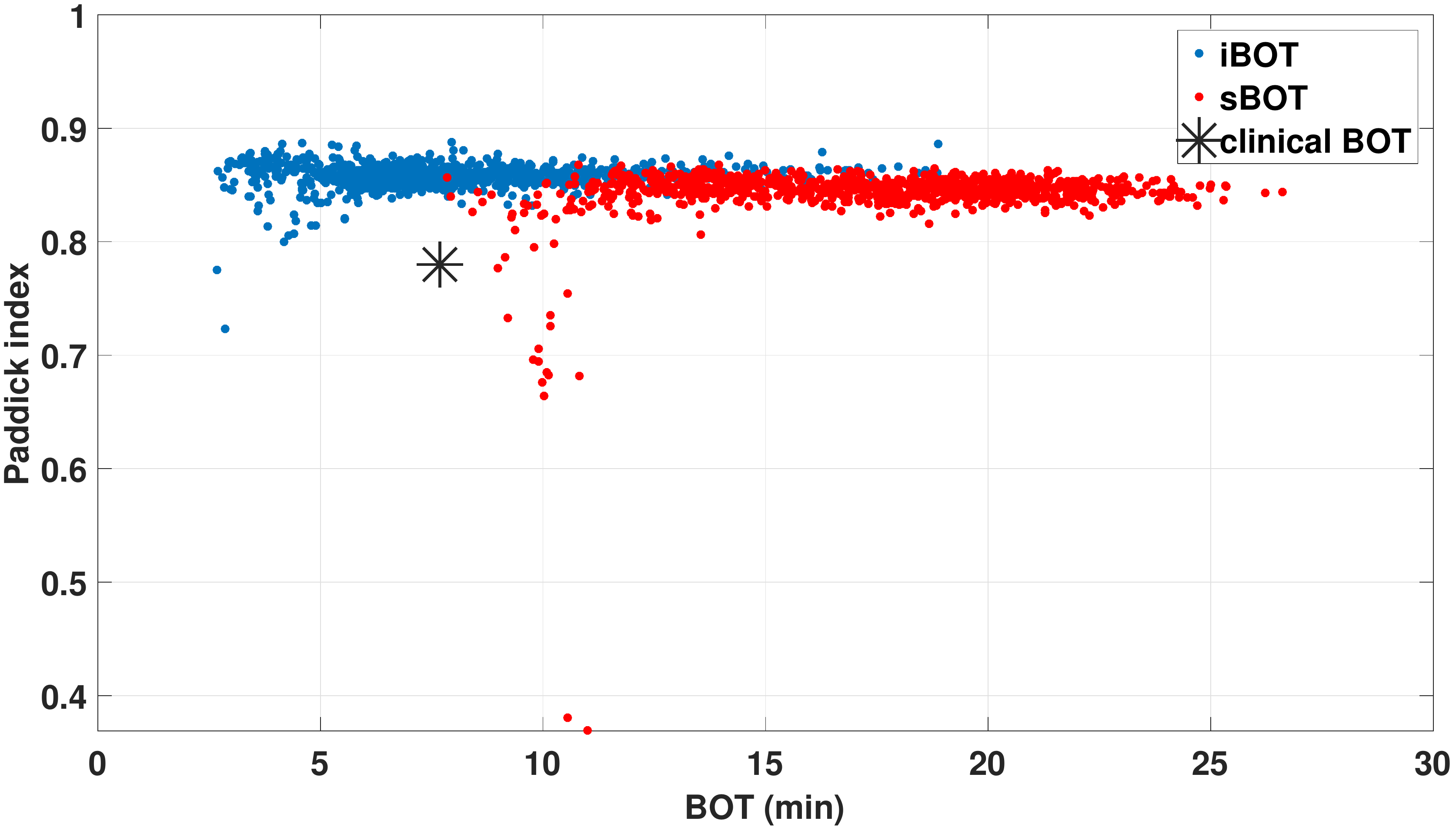}}%
 ~
 \subfigure[Gradient index vs. BOT]{%
 \label{fig:GIvsBOT}%
\includegraphics[width = 0.5\textwidth]{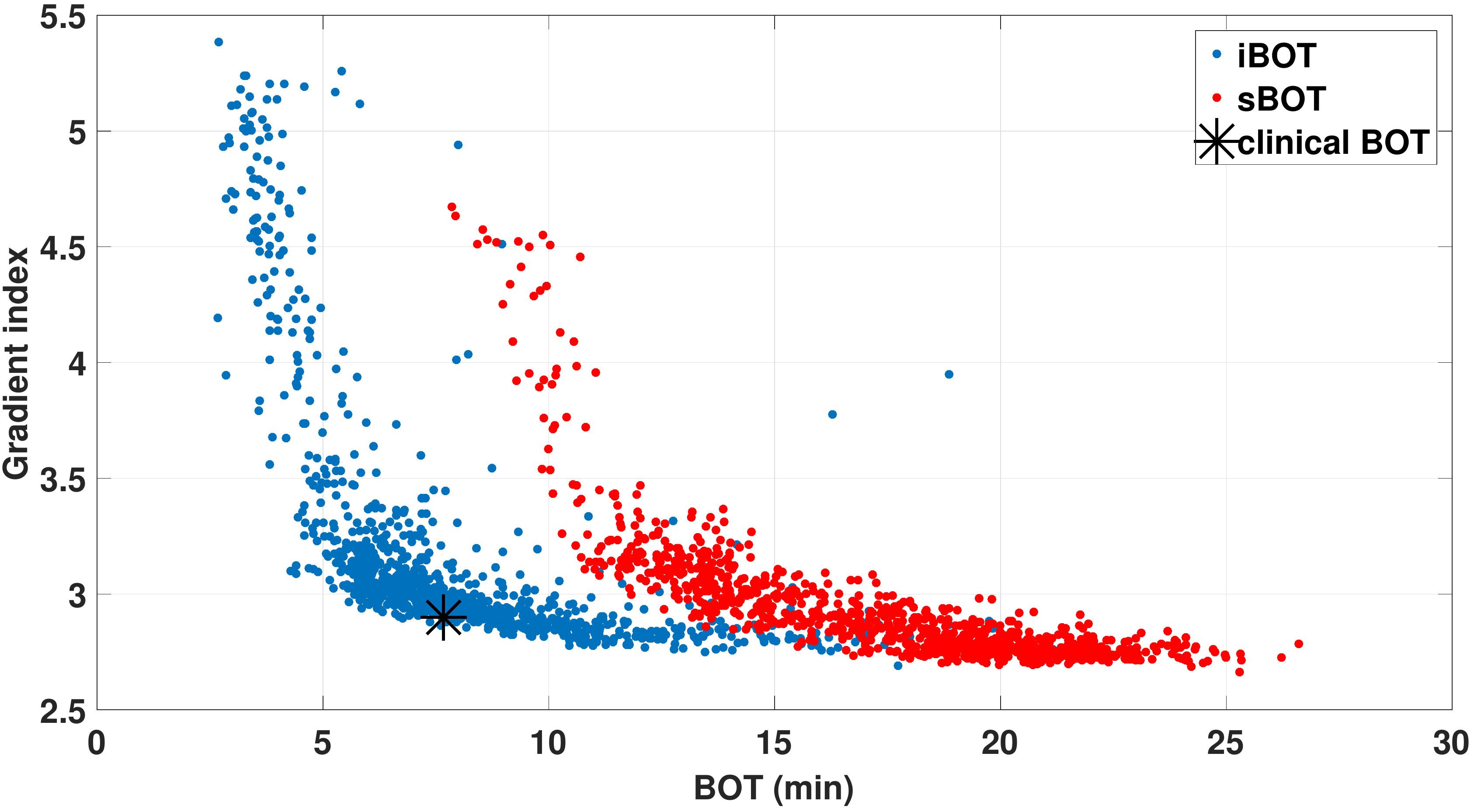}}%
 \caption{Comparison of the two BOT penalization terms for the small acoustic neuroma case.}\label{fig:iBOT_vs_BOT2}
 \end{figure}

\begin{figure}%
 \centering
 \subfigure{%
 \includegraphics[width = 0.3\textwidth]{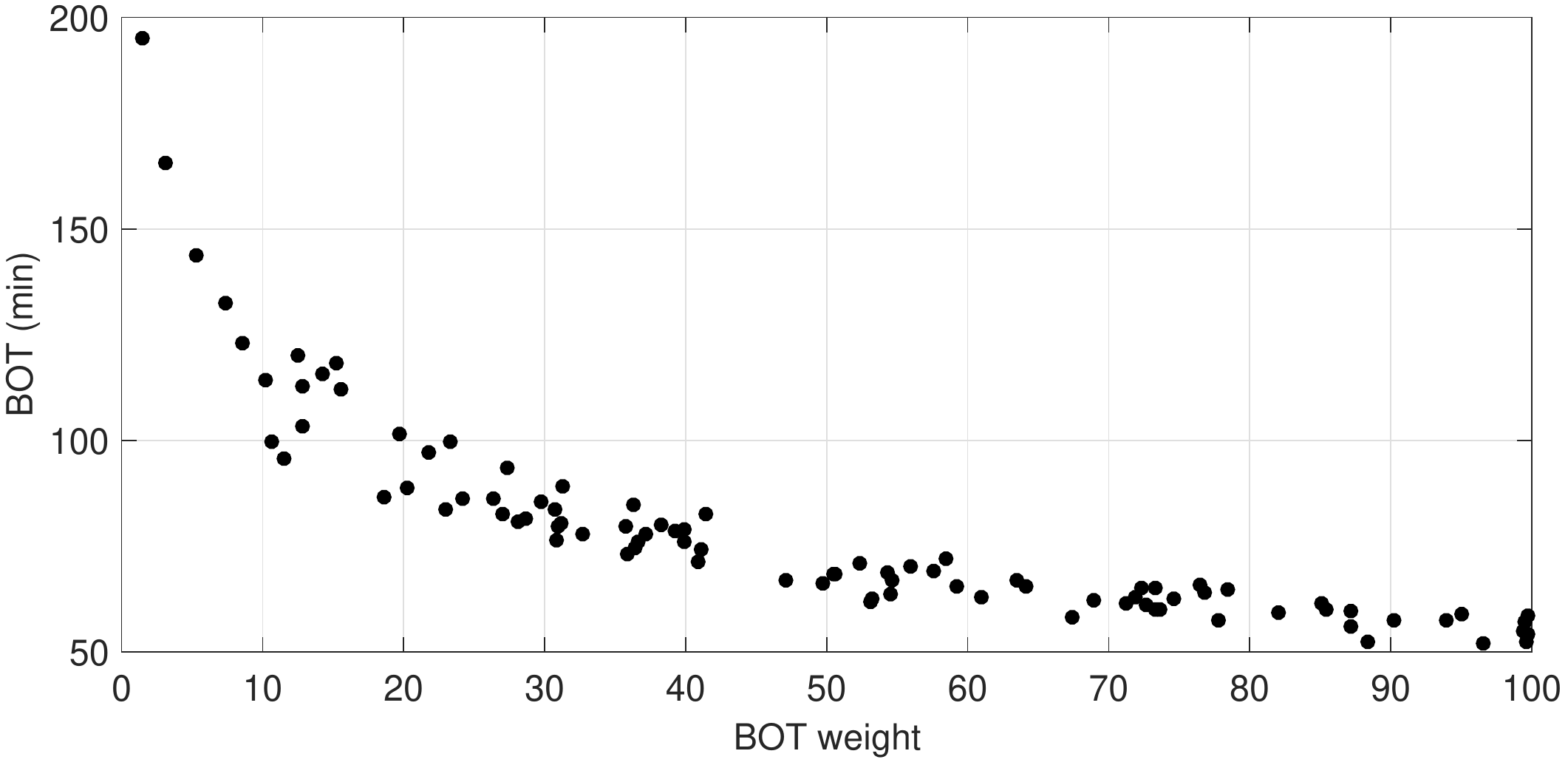}}%
 \subfigure{%
 \includegraphics[width = 0.3\textwidth]{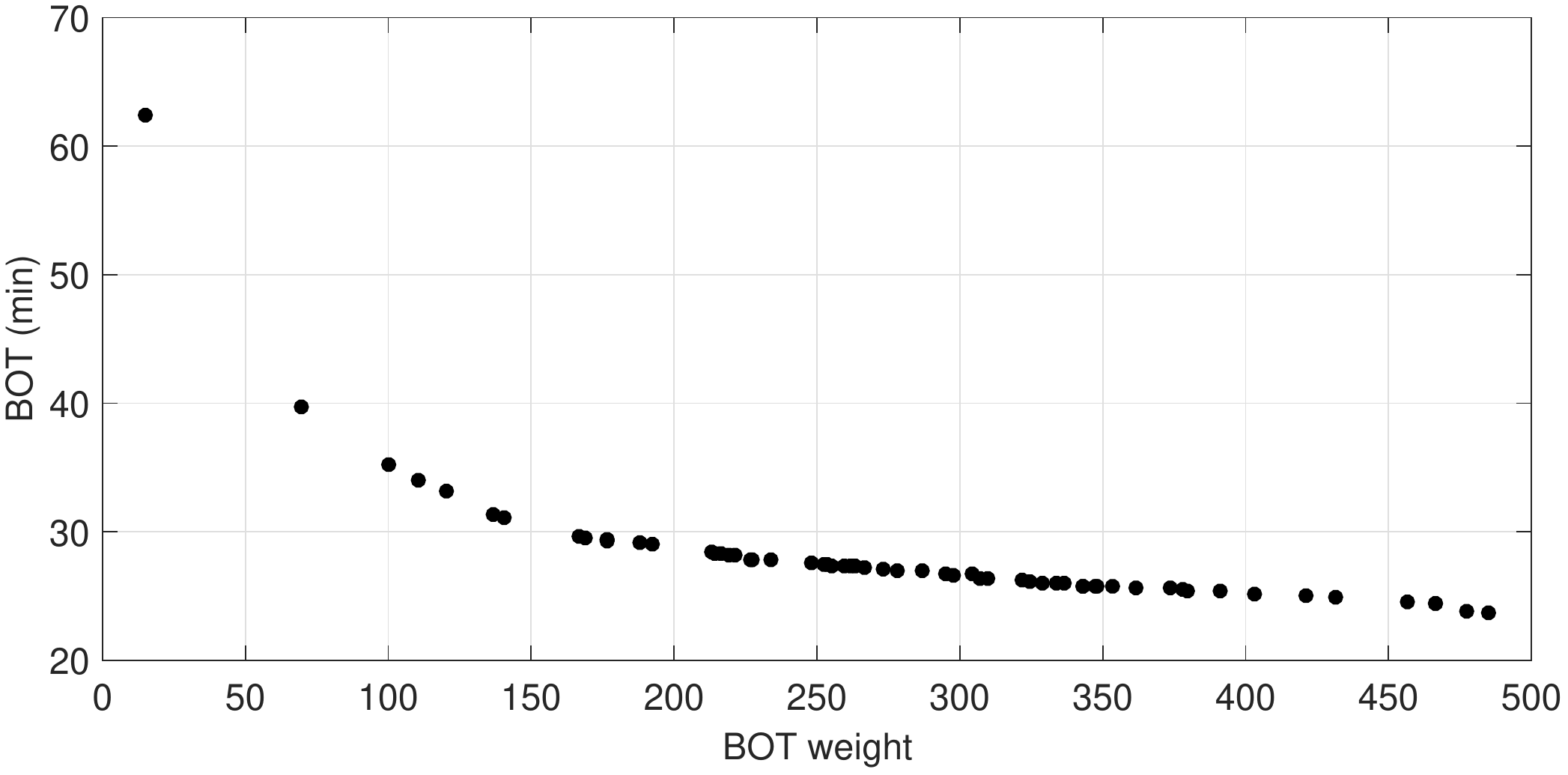}}%
 \subfigure{%
 \includegraphics[width = 0.3\textwidth]{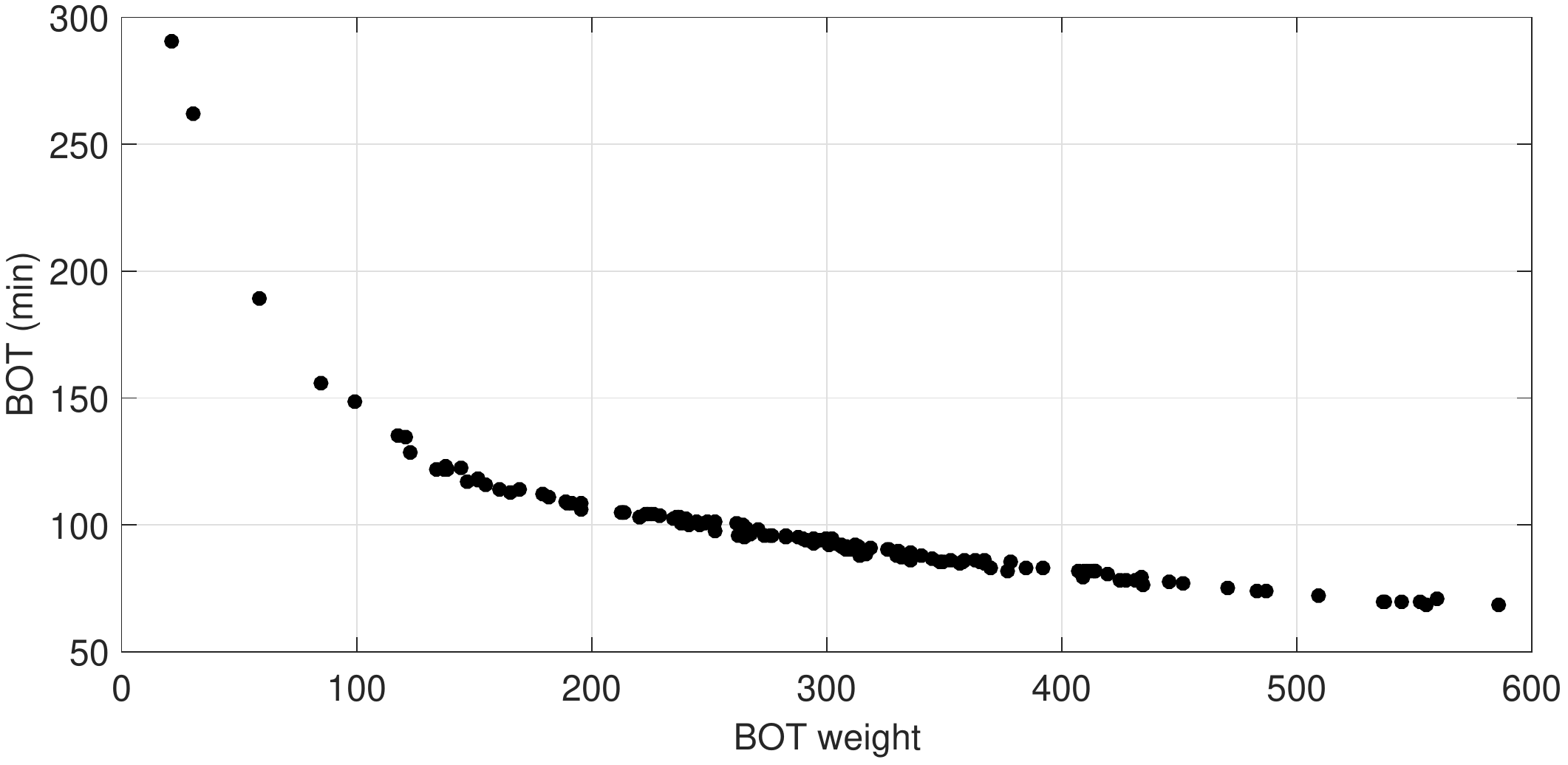}}%
 \\
 \subfigure{%
\includegraphics[width = 0.3\textwidth]{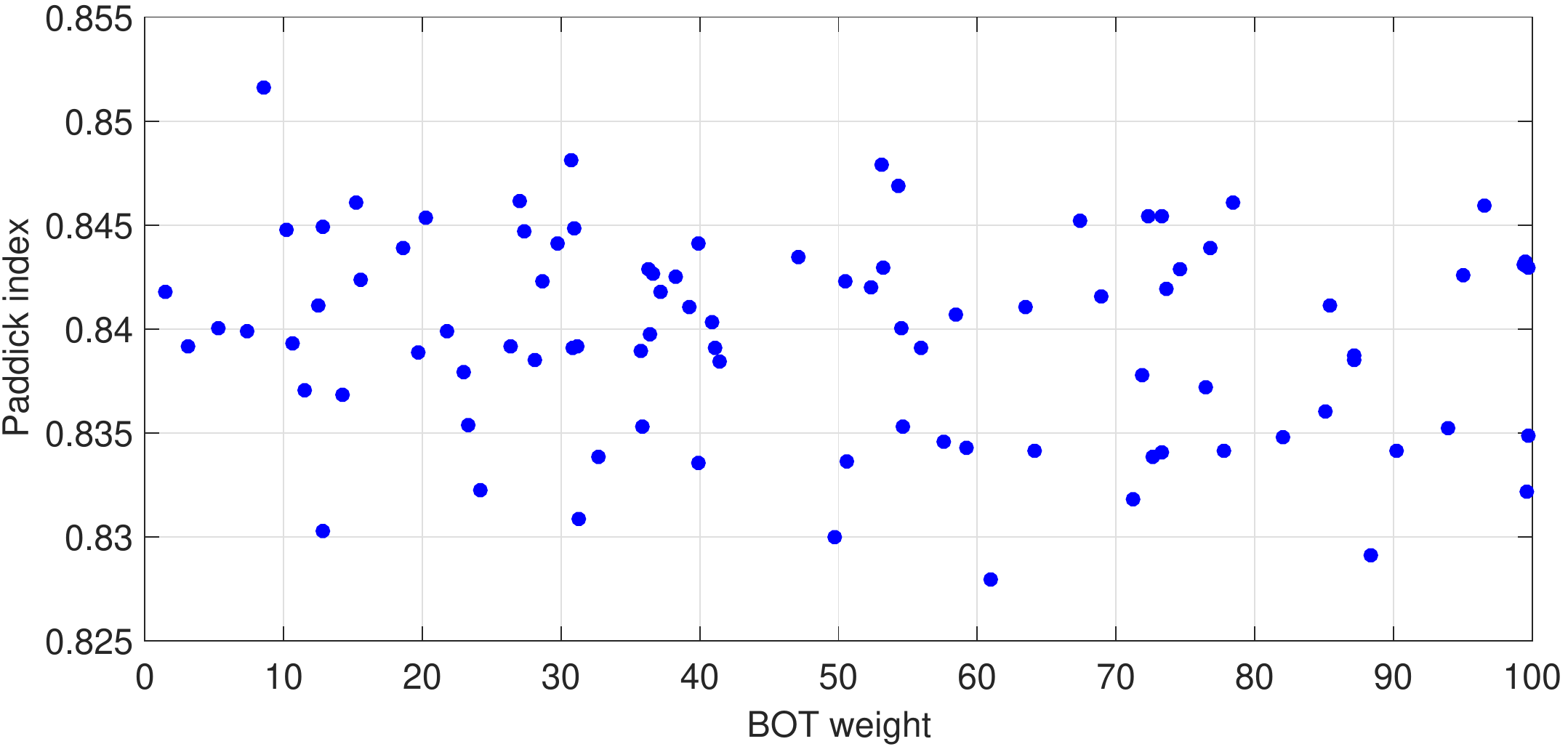}}%
\subfigure{%
\includegraphics[width = 0.3\textwidth]{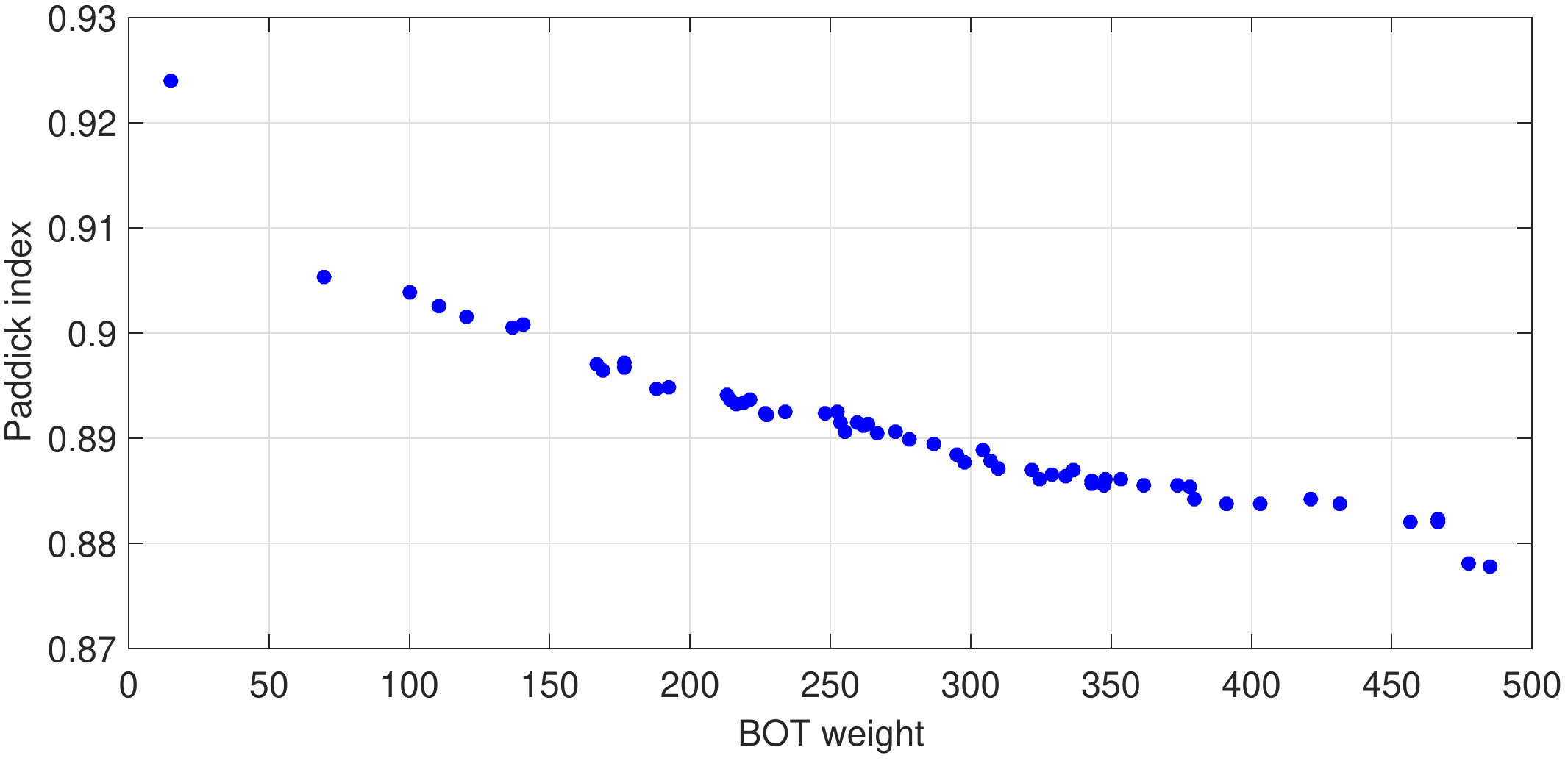}}%
\subfigure{%
\includegraphics[width = 0.3\textwidth]{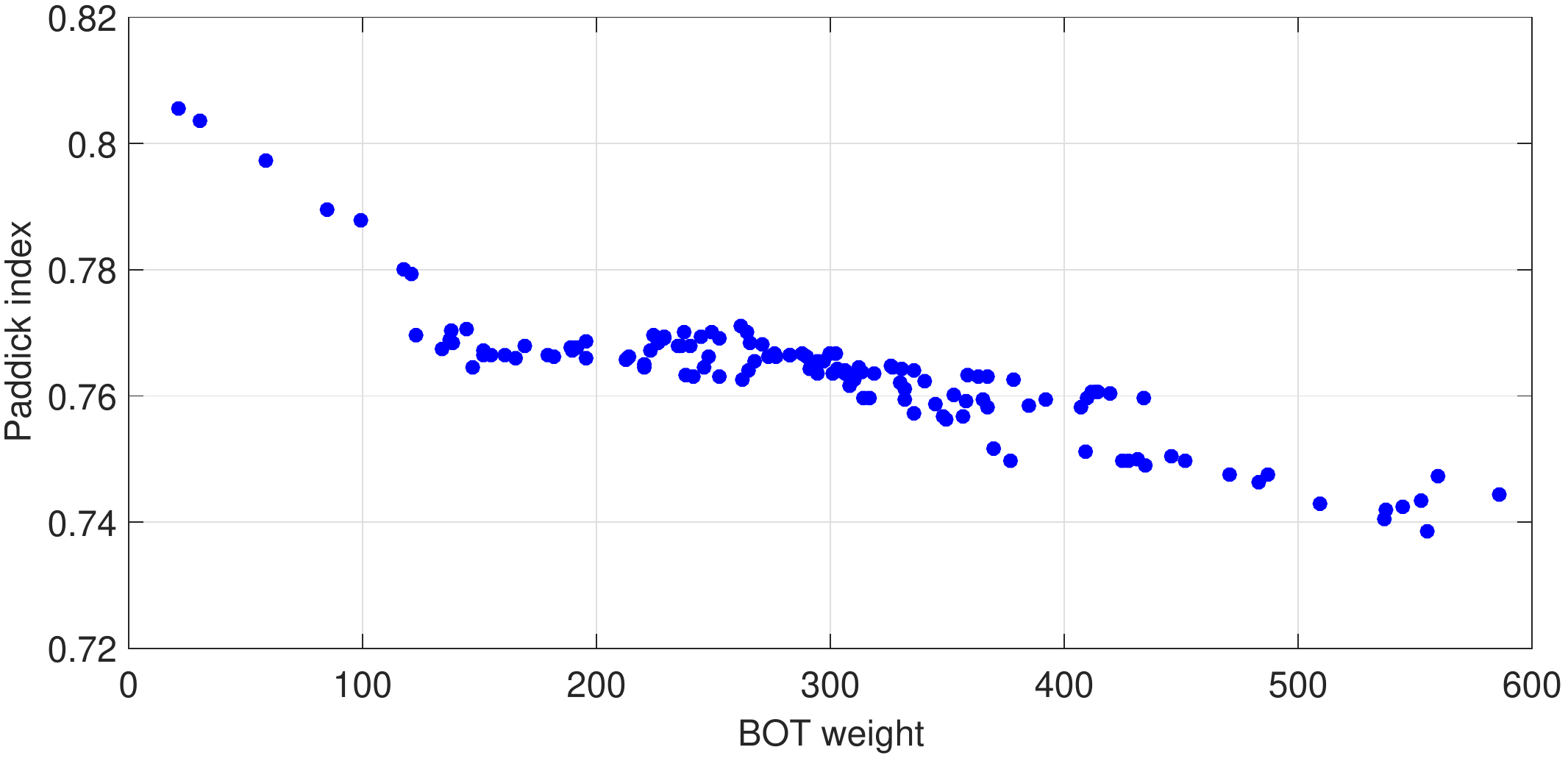}}%
\\
\subfigure{%
\includegraphics[width = 0.3\textwidth]{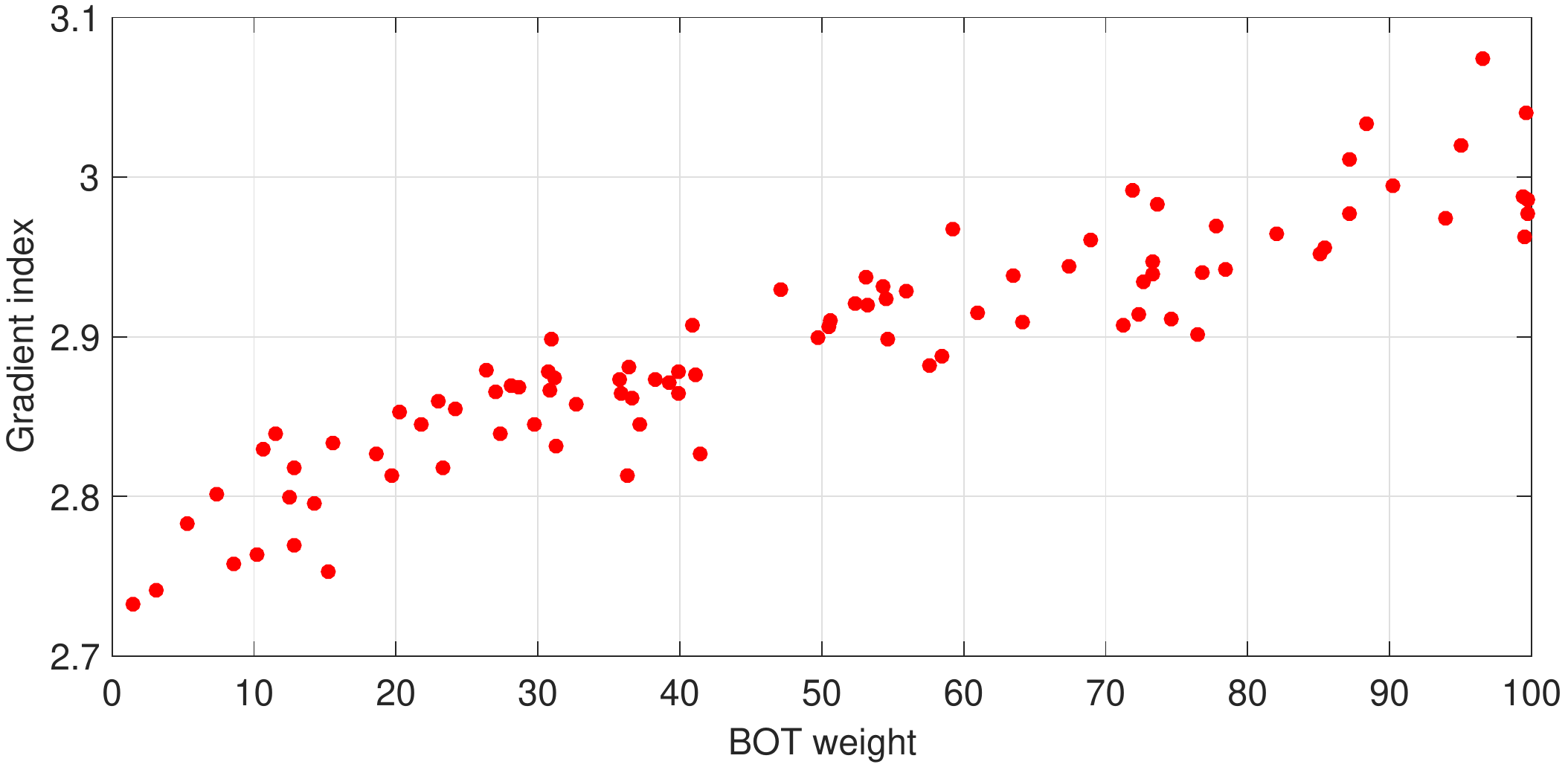}}%
\subfigure{%
\includegraphics[width = 0.3\textwidth]{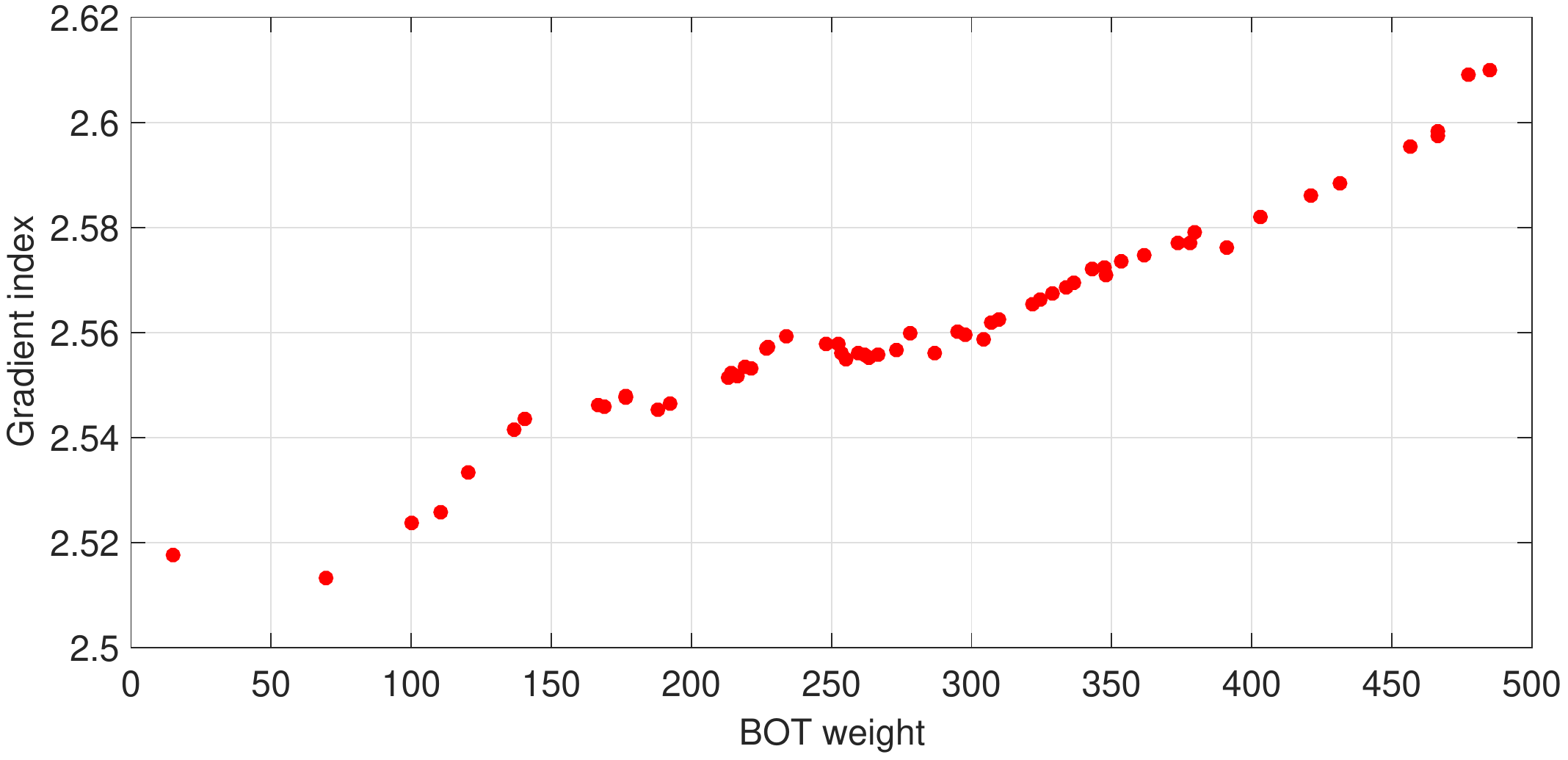}}%
\subfigure{%
\includegraphics[width = 0.3\textwidth]{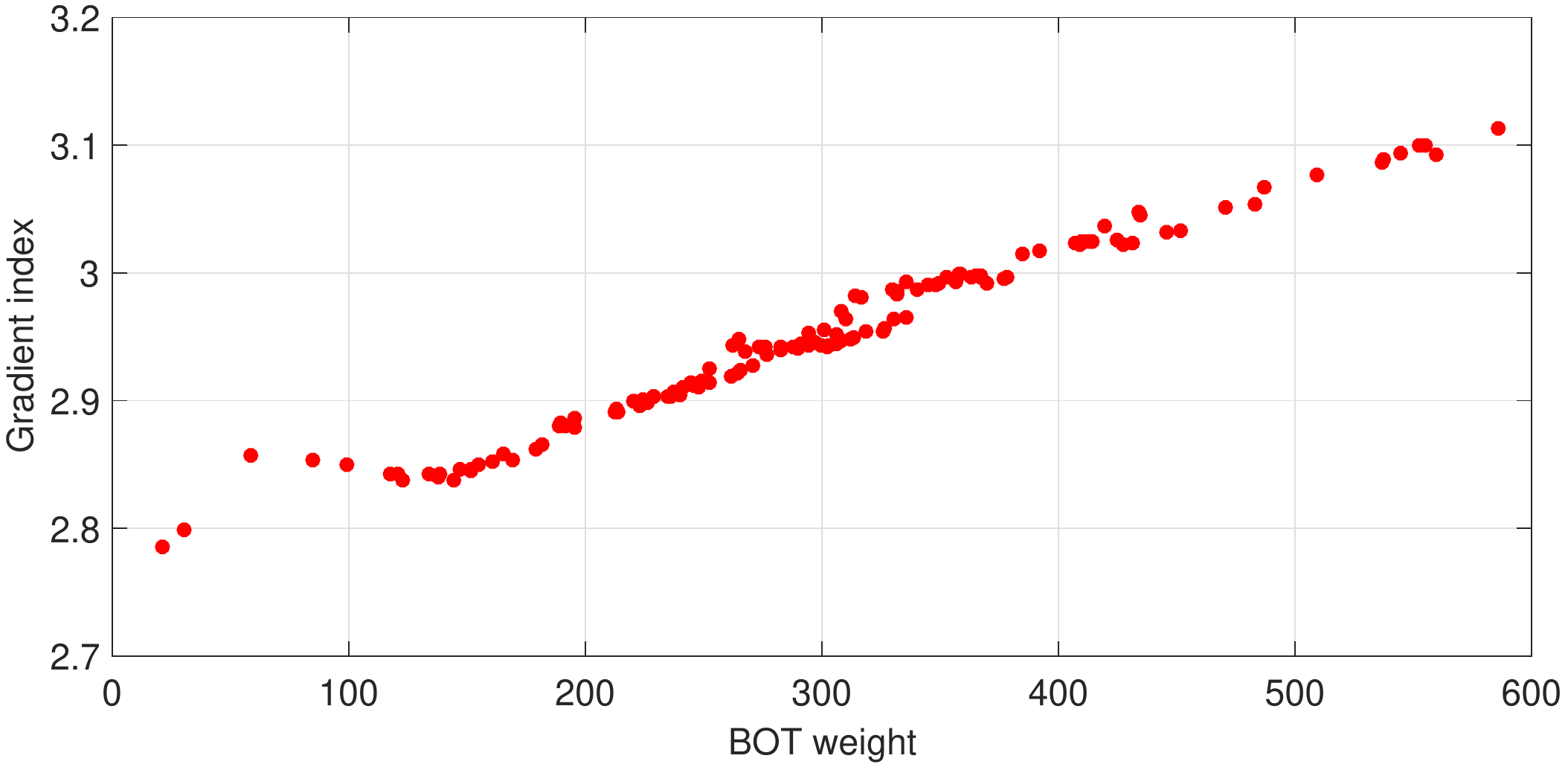}}%
 \caption{Dependence of BOT (top), Paddick index (middle) and gradient index (bottom) on the iBOT penalization. The columns correspond to different cases: small acoustic neuroma (left), medium acoustic neuroma (center) and irregular meningioma (right).}\label{fig:BOT_tradeoff}
 \end{figure}

To show the efficiency of the iBOT penalization, we compare with replacing the term $\Theta(t)$ in  Eq.~\eqref{eq:costfcn} with a simple summation over all times, 
\begin{equation}
\Theta_\text{simple}(t) = \sum_{i=1}^{N_\text{iso}}\sum_{s=1}^8\sum_{c=1}^3 t_{isc},
\end{equation}
which we refer to as simple BOT (sBOT). The rest of the objective function in  Eq.~\eqref{eq:costfcn} is left unchanged.  We make the comparison on three different cases: one small acoustic neuroma, one medium sized acoustic neuroma and one irregular meningioma. By varying the weight of the (selectivity promoting) inner shell against the BOT penalization while keeping the other weights fixed, we obtain plans with different plan quality and BOTs. To get adequate statistics, we ran 1000 optimizations for each case and choice of penalization.
Plans with the same Paddick index and gradient index to within $\pm$ 1 $\%$  were tallied and the average BOT calculated for each bin. The bins were chosen to have Paddick and gradient indices within $\pm$ 5 \% of the clinical plan. Table \ref{tab:sBOT} summarizes the mean and standard deviation of the average BOT with the iBOT penalization and the average BOT with the sBOT penalization, as well as the mean and standard deviation of the ratios computed for each bin. 
\begin{center}
\renewcommand\tabcolsep{6pt}
\begin{table}[ht]
    \caption{BOT (min) at 3 Gy/min when using different penalization terms}\label{tab:sBOT}
    \begin{tabular}{ l  c  c  c  }
    \hline\hline
     & Small acoustic neuroma & Medium acoustic neuroma & Irregular meningioma \\ \hline
    iBOT& 9.3 (2.6) & 21.6 (1.5) & 110.8 (3.1)  \\ 
    sBOT & 16.8 (1.9) & 75.2 (4.4) & 303.2 (11.6) \\ 
    iBOT/sBOT & 0.55 (0.16) & 0.29 (0.026) & 0.37 (0.029) \\ 
    Optimization times iBOT (s) & 0.135 & 3.2 & 3.3 \\ 
    Optimization times sBOT (s) & 0.125 & 3.4 & 3.5 \\ 
    \hline\hline
    \end{tabular}
    \end{table}
\end{center}
Clearly, iBOT results in plans of equal quality but markedly shorter BOT than sBOT, in particular for the two larger targets. The optimization times are practically equal.
 Figure~\ref{fig:iBOT_vs_BOT2} gives an overview of all the simulations for the small acoustic neuroma, for which the BOT reduction was the smallest (but still almost a factor of 2). 
Figure \ref{fig:BOT_tradeoff} shows how the BOT, Paddick index and gradient index depend on the iBOT weight. Even though the Paddick index may appear independent of the iBOT in the small acoustic neuroma case (middle left), recall from Eq. \eqref{eq:PI} that it is the product of coverage and selectivity. Separately these are not constant as a function of BOT; a shorter BOT is obtained by increasing the total dose rate, which for Gamma Knife is done by using larger collimators and more sectors simultaneously. This smoothens the dose distribution so that coverage increases at the expense of selectivity and gradient index.

\subsection{Subsampling}
Since we subsample stochastically, the resulting plan metrics also become stochastic. We do, however, want to choose the subsampling fractions such that the statistical fluctuations remain reasonably small. Here, we require the standard deviation of both coverage and selectivity to be below 1\%. 
Furthermore, we want to ensure that our sampling strategy, which samples both interior and surface points, performs at least comparably to sampling only in the interior.

\begin{figure}[ht!]%
\centering
\subfigure[Coverage]{%
\label{fig:first}%
\includegraphics[width = 0.47\textwidth]{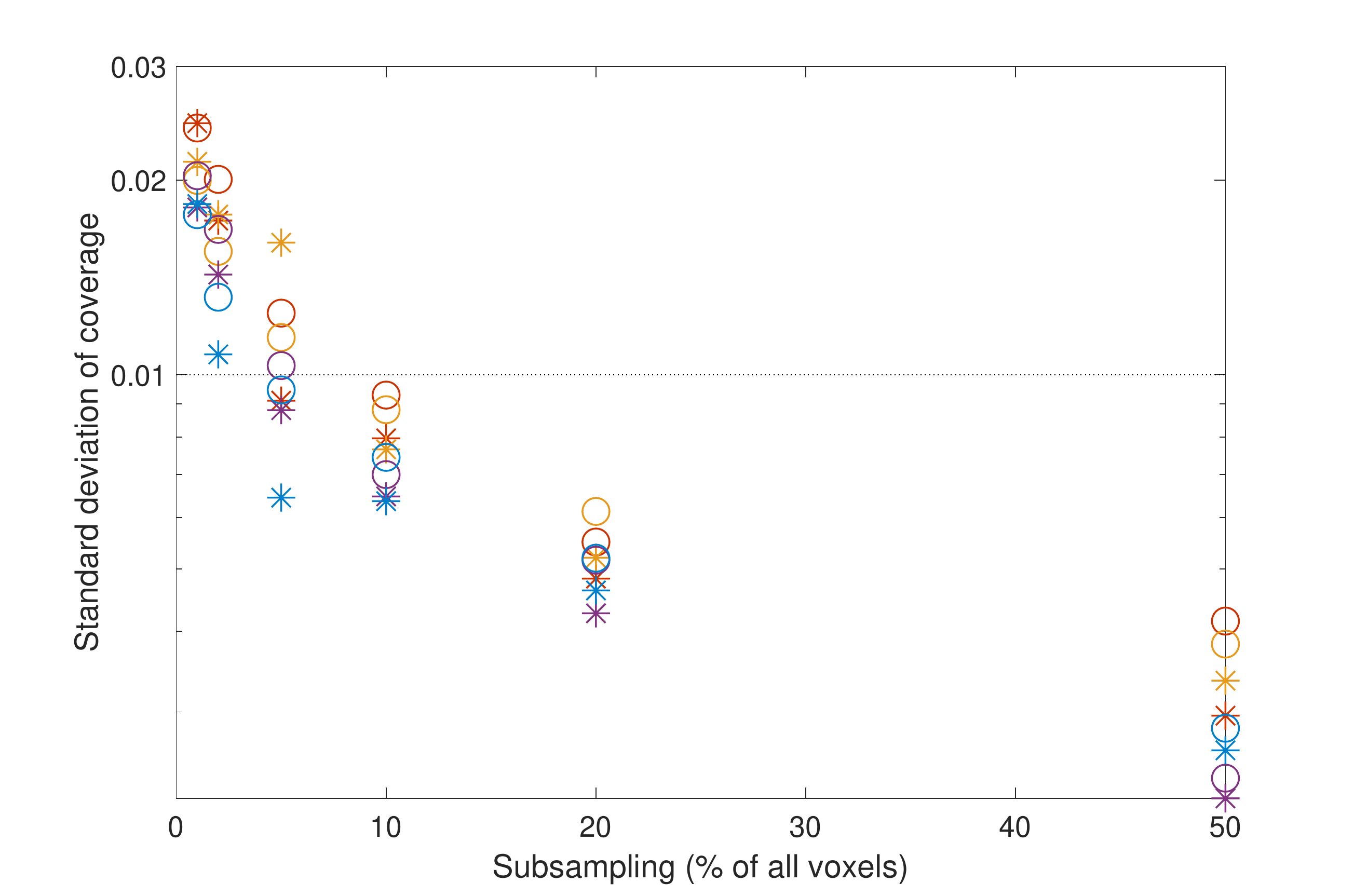}}%
~
\subfigure[Selectivity]{%
\label{fig:second}%
\includegraphics[width = 0.5\textwidth]{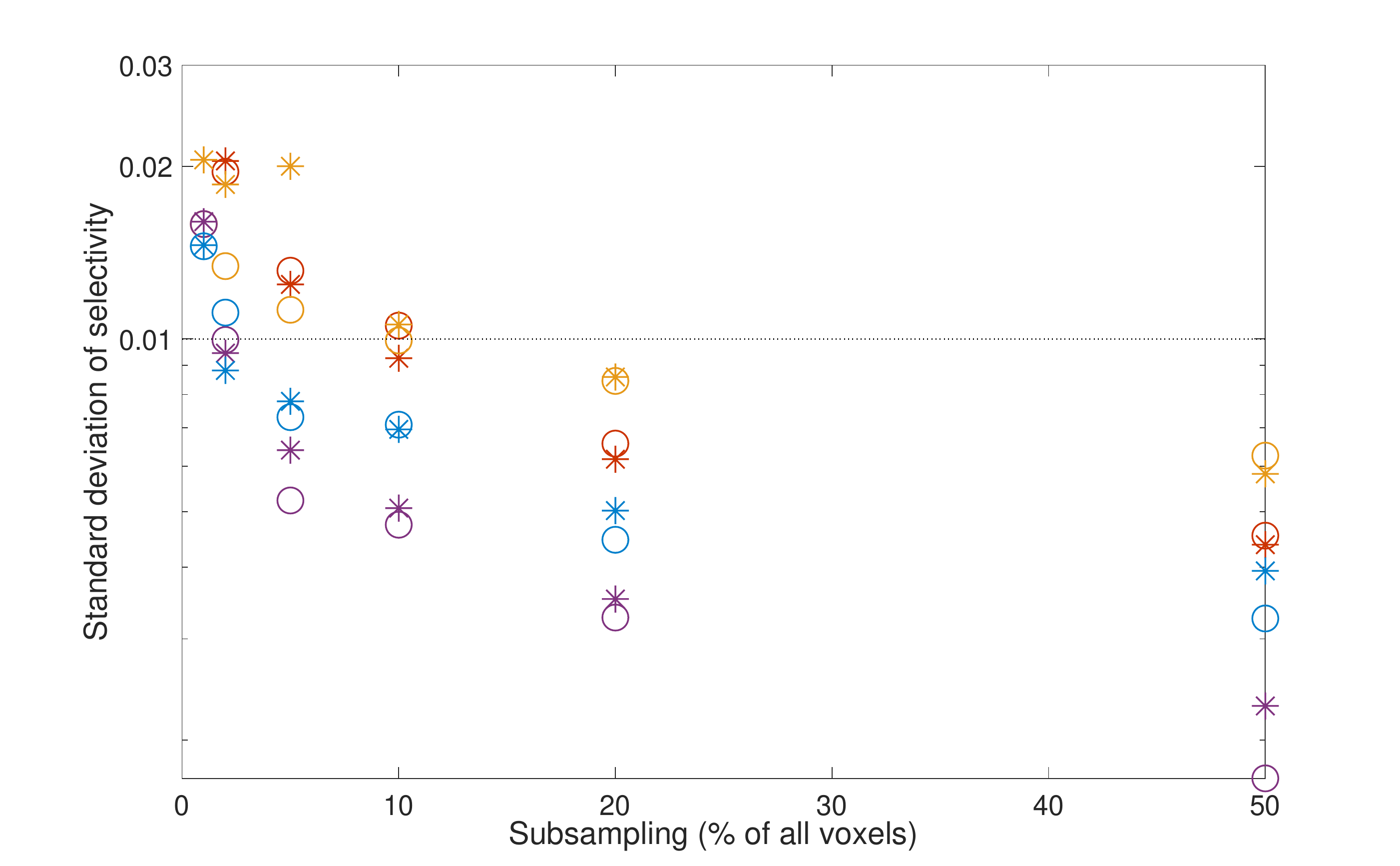}}%
\caption{The standard deviation of coverage and selectivity for the two sampling methods, interior and surface ($*$) and interior only ($\circ$) for the small acoustic neuroma (red), the medium acoustic neuroma (purple), the irregular meningioma (yellow) and the arteriovenous malformation (blue).}\label{fig:sampling}
\end{figure}

We compare the two strategies on the three cases studied in Sec.~\ref{sec:BOT} and an arteriovenous malformation case (AVM). Before sampling, the number of target voxels are 6119 (small acoustic neuroma), 12078 (medium acoustic neuroma), 27765 (meningioma) and 30326 (AVM). We adjust the voxel size for the interior point sampling so that the total number of points in the target is the same for both sampling methods.  
We performed 100 runs for each of seven subsampling fractions. Figure~\ref{fig:sampling} presents the resulting standard deviation of coverage and selectivity as a function of the percentage of the total number of voxels, i.e. the voxels belonging to all structures in the problem. For each sampling method we used a fixed weight setting ($(w_{\rm T},w_{\rm S},w_{\rm G},w_{\rm BOT})=(1.0,0.15, 0.15, 0.15)$ for our sampling method and $(w_{\rm T},w_{\rm S},w_{\rm G},w_{\rm BOT})=(1.0,0.025, 0.025, 0.25)$ for interior sampling only) that rendered clinically acceptable plans with metrics within 1\% of each other.
From Fig.~\ref{fig:sampling}, we conclude that our sampling strategy performs  comparably to the interior point sampling for both coverage and selectivity. Also, we see that the standard deviation of both metrics is below 1\% when the subsampling fraction is at least 10\%. The value of the metrics, and thus the standard deviation, is computed on the original grid. For small sampling fractions, the mean will deviate from the true value. However, for the sampling sizes we are interested in, i.e. about 10\%, this is typically not an issue.

\begin{figure}
\includegraphics[width = 0.6\textwidth]{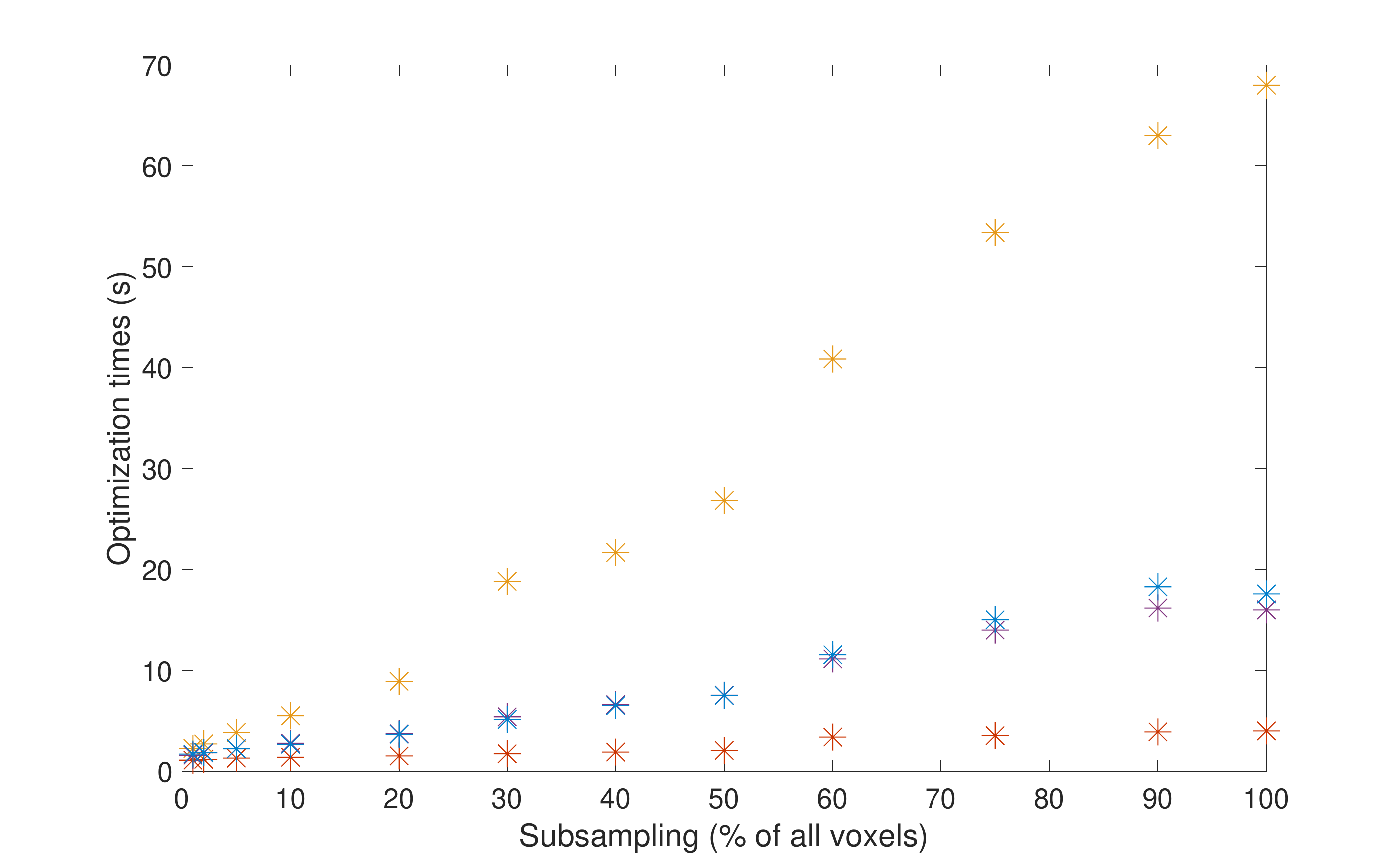}
\caption{The optimization time as a function of subsampling fraction for small acoustic neuroma (red), the medium acoustic neuroma(purple), the irregular meningioma (yellow) and the arteriovenous malformation (blue).}\label{fig:opt_times}
\end{figure}

As illustrated in Fig.~\ref{fig:opt_times}, subsampling shortens the optimization time. Evidently, the dependence between the optimization time and the subsampling fraction is approximately linear. In the cases considered above, asubsampling of 10\% shortens the optimization time by a factor 8--22 compared to using all initial sampling points.  
\subsection{Overall performance}
\begin{figure}%
\centering
\subfigure[Selectivity for all the cases]{%
\includegraphics[width = 0.65\textwidth]{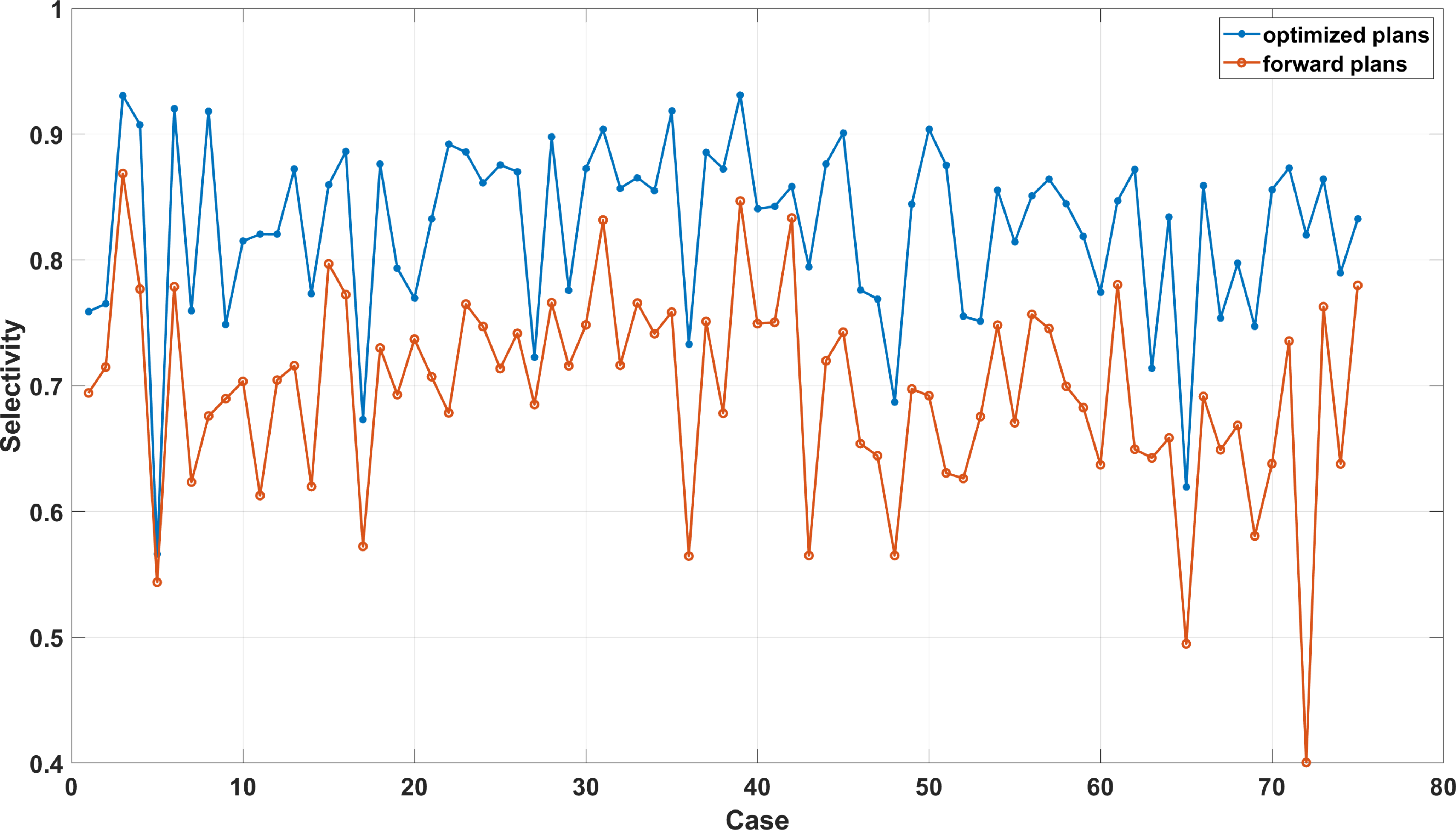}}%
\\
\subfigure[BOT for all the cases]{%
\includegraphics[width = 0.65\textwidth]{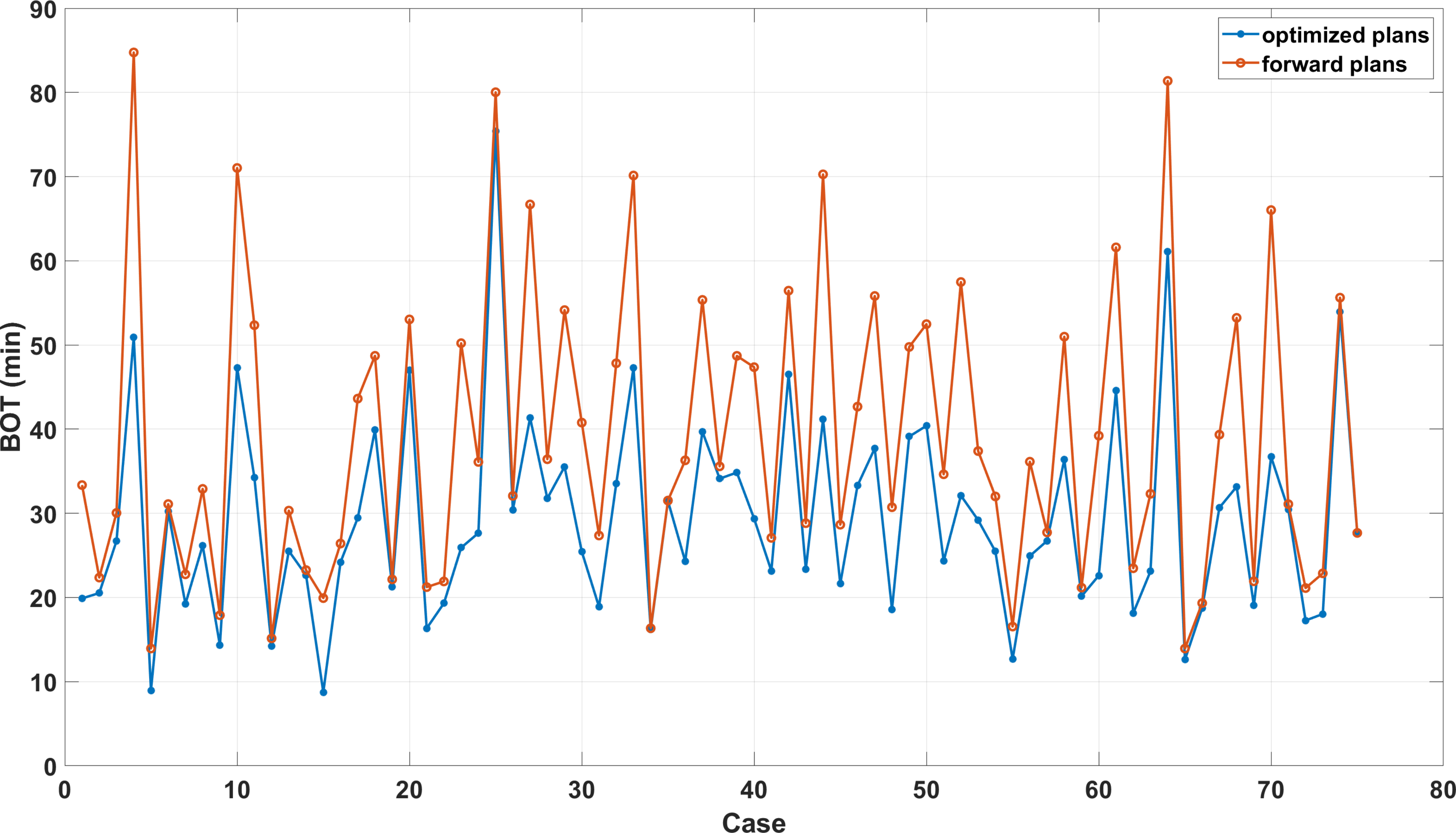}}%
\\
\subfigure[Gradient index for all the cases]{%
\includegraphics[width = 0.65\textwidth]{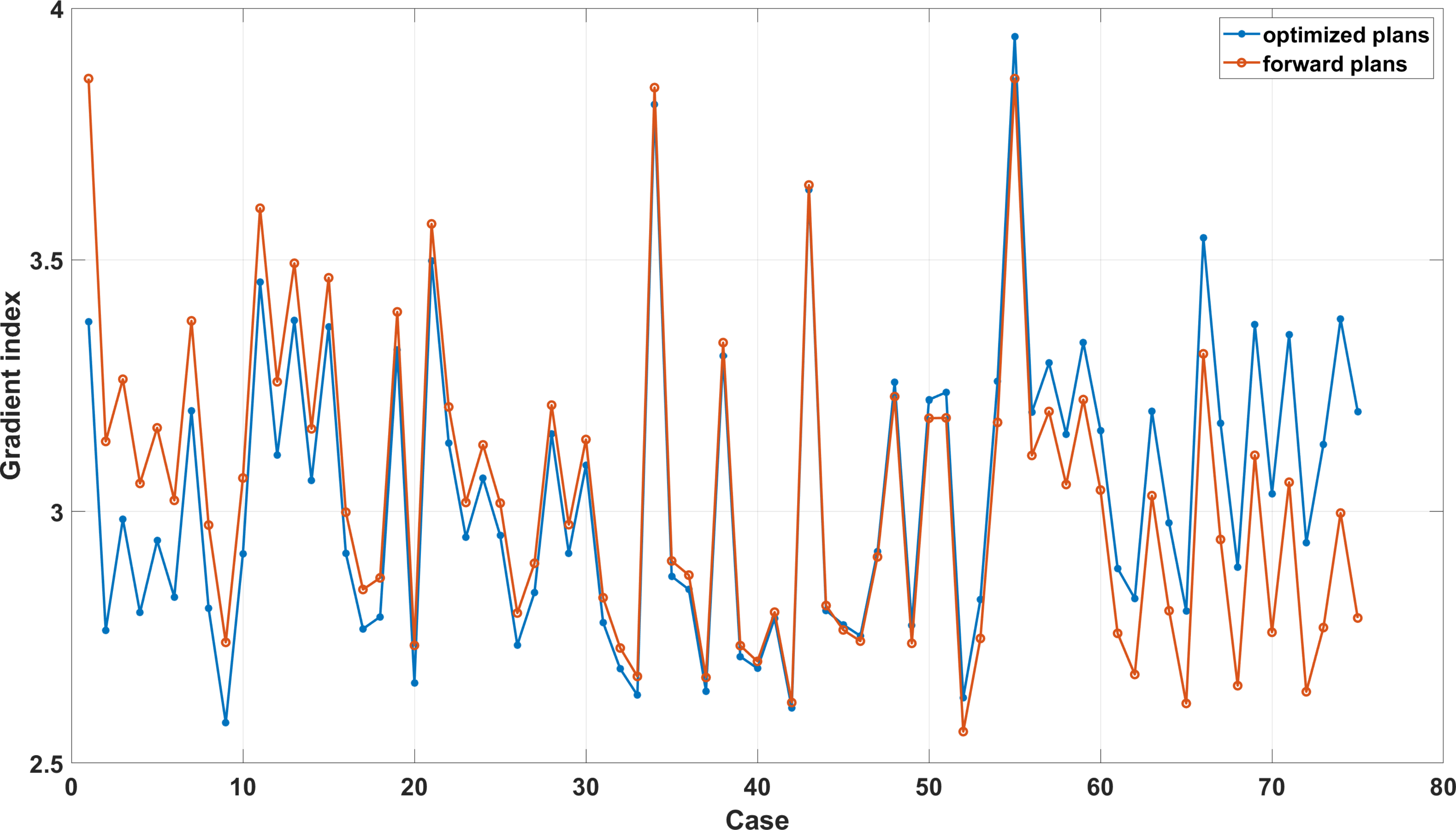}}%
\caption{Metrics for the optimized plans compared to the forward plans. The cases are sorted by the difference in gradient index.}\label{fig:75_cases_figure}
\end{figure}

We evaluate the overall performance by comparing the plan metrics for optimized plans with manual forward plans for 75 clinical cases (the majority of Gamma Knife users still do forward planning). The clinical cases are all single target cases: single metastases, acoustic neuromas and meningiomas. The range in tumor size is 0.6--11.7 ${\rm cm}^3$ and the planning dose range is 12--24 Gy.  In 18 of these cases there are organs at risk (e.g. brainstem, cochlea, optic nerves) so, to make a fair comparison, the maximum doses to the OARs in the manual plans are used as hard constraints in the optimization. Furthermore, all the manual plans have 98--100\% coverage, so the weights were chosen to always satisfy this criteria. To compare the manual and optimized plans, we have to find weight settings for each optimized plan that give a similar trade-off between clinical objectives as the corresponding manual plan. For simplicity, we employed the following strategy: first, we created a range of plans with different characteristics by performing 101 optimizations for each case, with randomly sampled weights for the inner ring and BOT penalization; then we selected, for each case, the optimized plan with the best gradient index among those with both higher selectivity and shorter BOT than the corresponding manual plan. The resulting plan metrics are shown in figure~\ref{fig:75_cases_figure}. In summary, for all of the 75 cases we could find optimized plans with simultaneously higher selectivity and shorter BOT than the manual plans, and in 44 cases these plans also had better gradient index than the manual plans. In other words, we found plans that dominate the manual ones in almost 60\% of the cases. 

\begin{figure}[ht]
\centering
\includegraphics[scale=0.3]{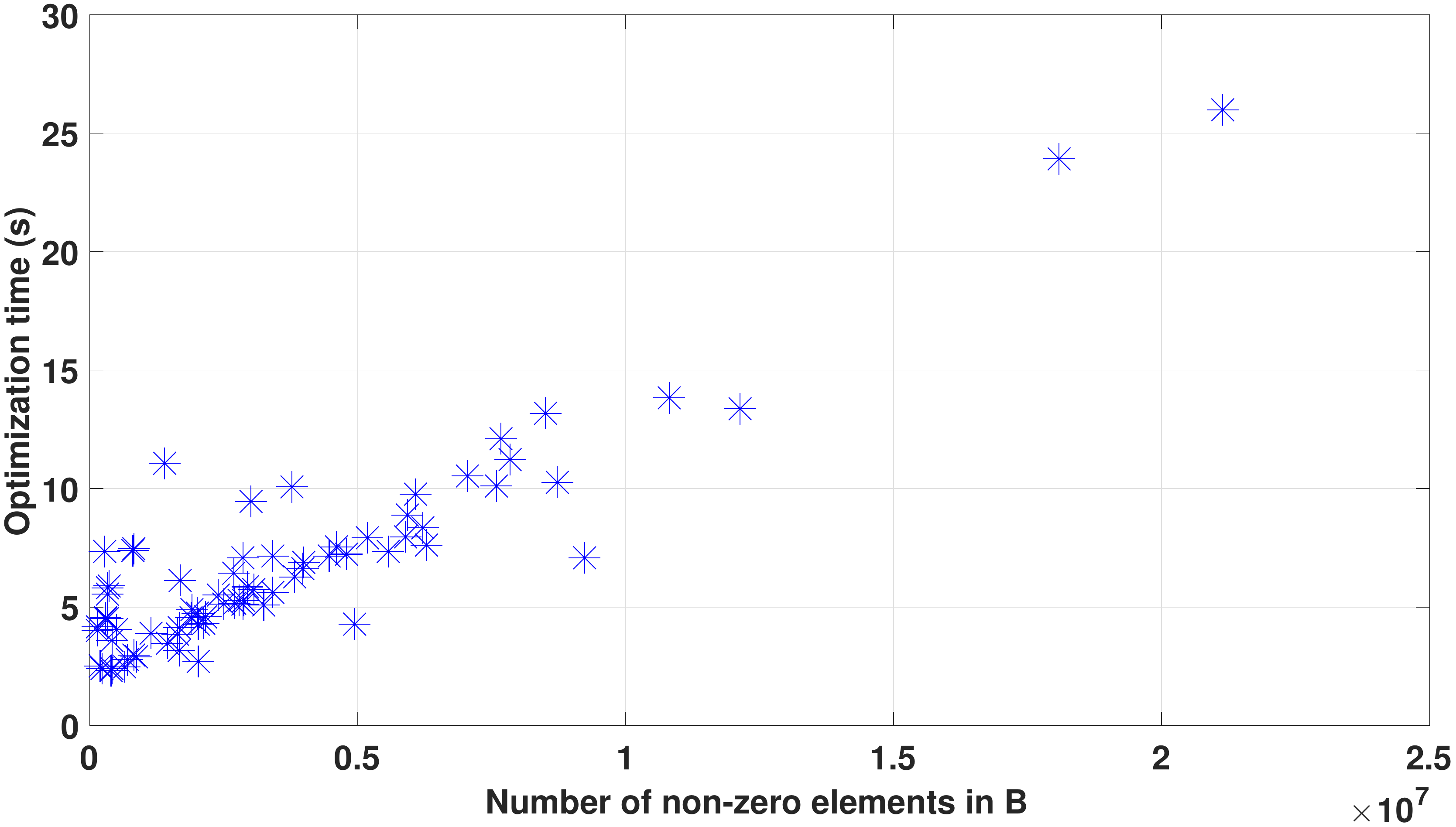}
\caption{Optimization time as a function of the number of non-zero elements in B for the 75 clinical cases. The times shown are averages over the 101 runs made for each case.}\label{fig:75_cases_sizeB_NZ}
\end{figure}

To solve the optimization problems, we used the open-source solver Glop\cite{Glop} with default settings. The optimization times ranged from 2.3 to 26 s with a median time of 5.7 s on a standard GammaPlan workstation (a HP Z640 with 32 Gb RAM and 12 cores running at 2.9 MHz). Figure~\ref{fig:75_cases_sizeB_NZ} shows how the optimization time depends on the number of non-zero elements in the constraint matrix $B$ (cf. appendix \ref{appendix:explicit_dual}). We conclude that there is a linear trend, but that the variation appears larger for small cases. 

\subsection{Detailed case study}\label{sec:GFRAME41}
\begin{figure}%
\centering
\subfigure[Reference]{%
\includegraphics[width = 0.33\textwidth]{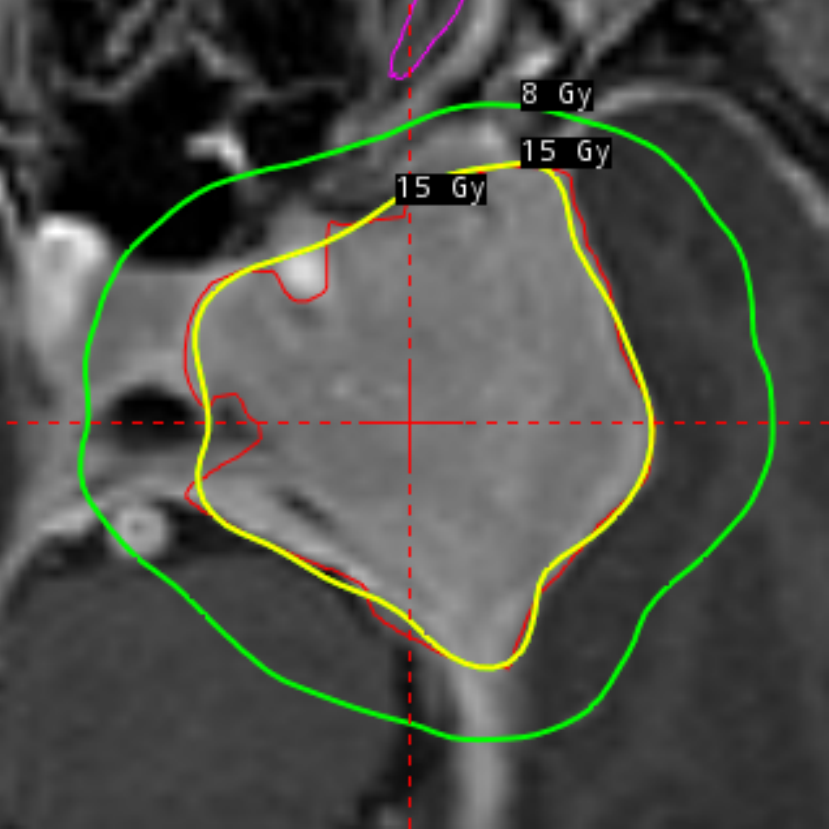}}%
\qquad
\subfigure[Reference]{%
\includegraphics[width = 0.55\textwidth]{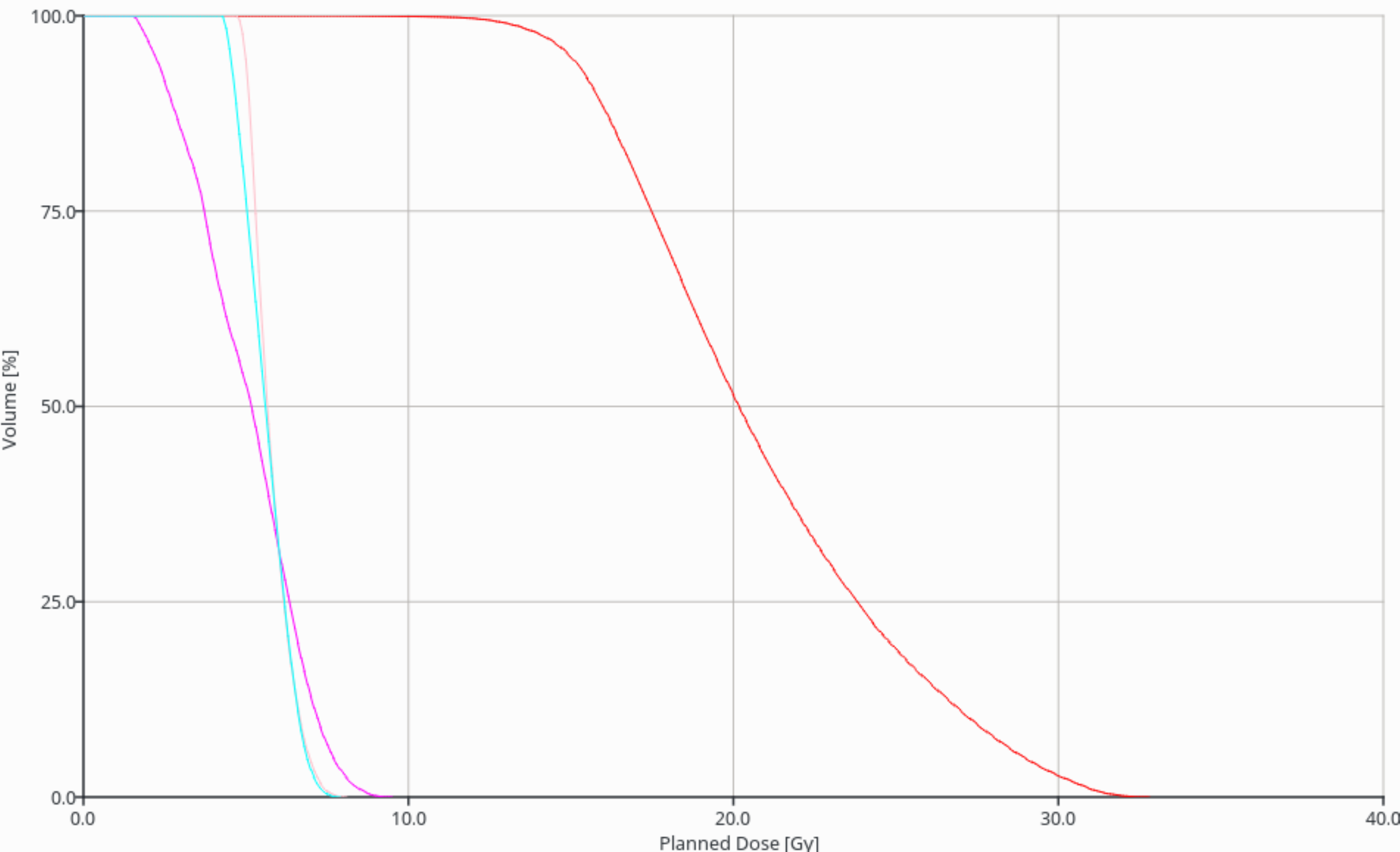}}%
\qquad
\subfigure[Promoting selectivity]{%
\label{fig:first}%
\includegraphics[width = 0.33\textwidth]{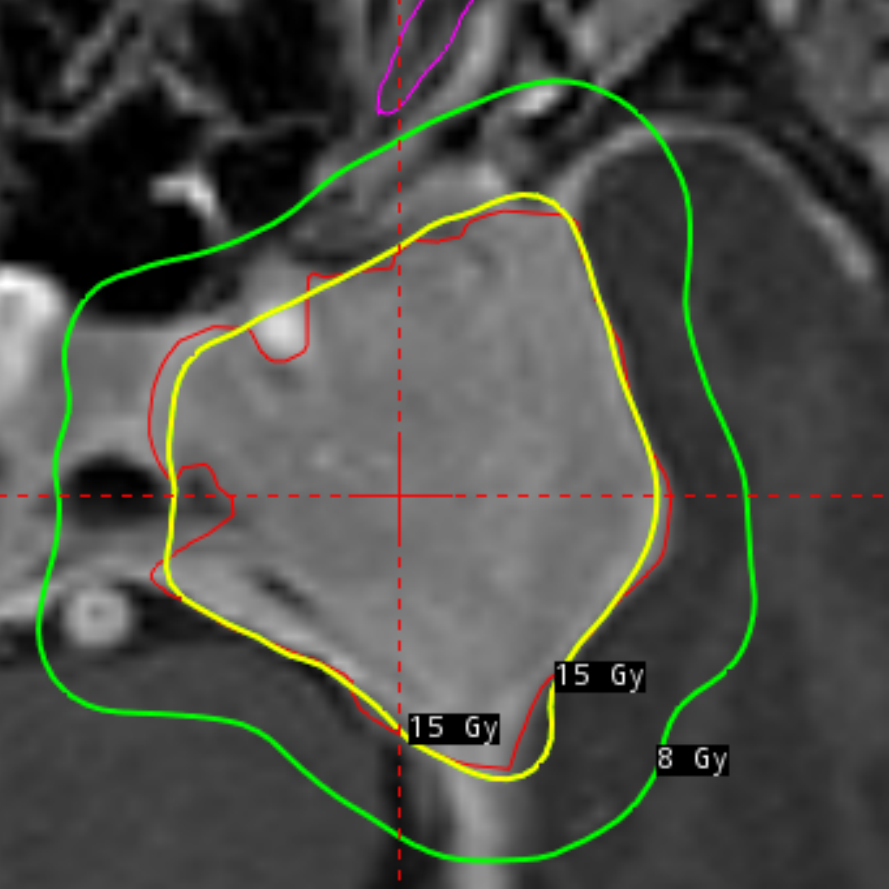}}%
\qquad
\subfigure[Promoting selectivity]{%
\includegraphics[width = 0.55\textwidth]{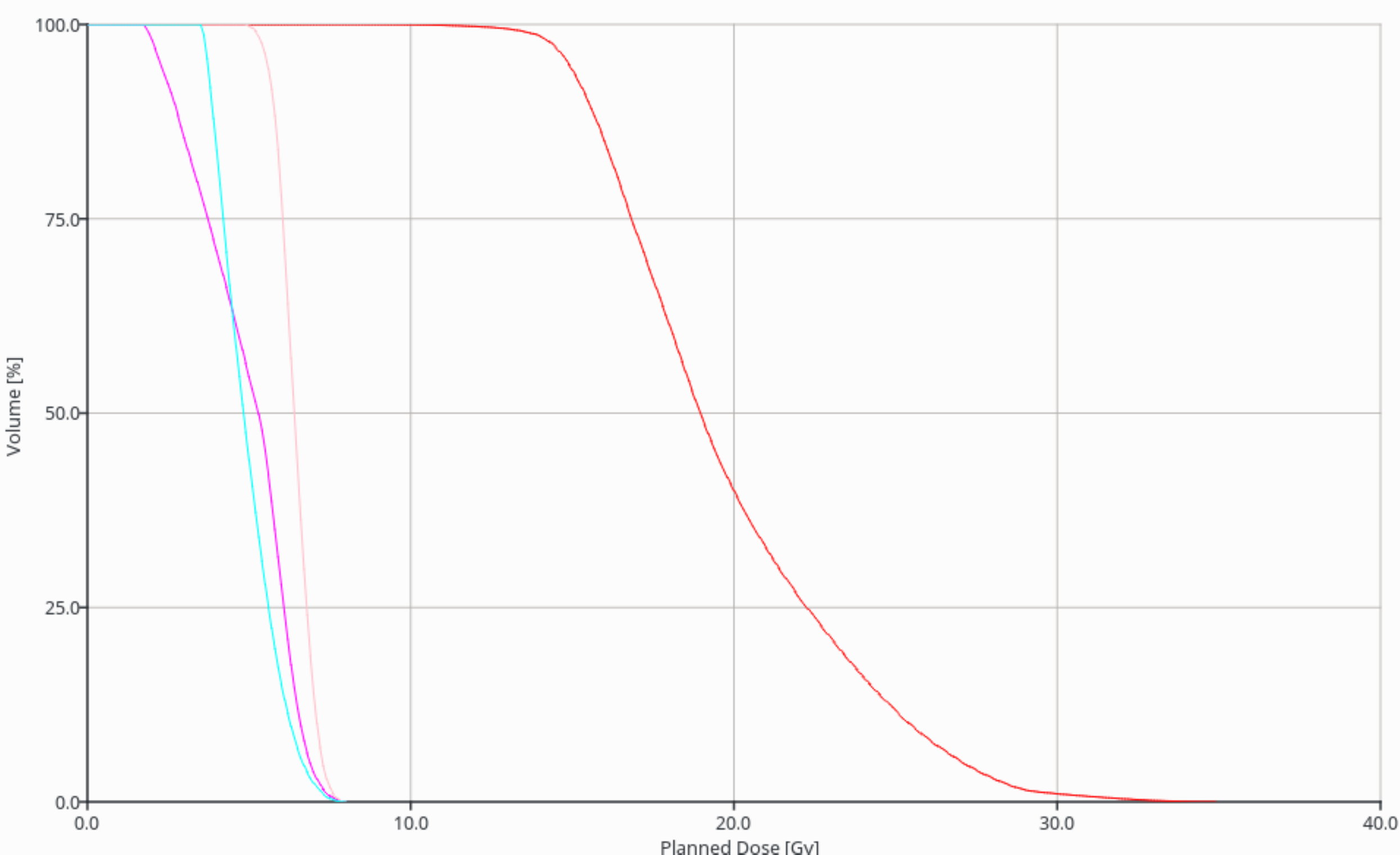}}%
\qquad
\centering
\subfigure[Promoting BOT]{%
\label{fig:first}%
\includegraphics[width = 0.33\textwidth]{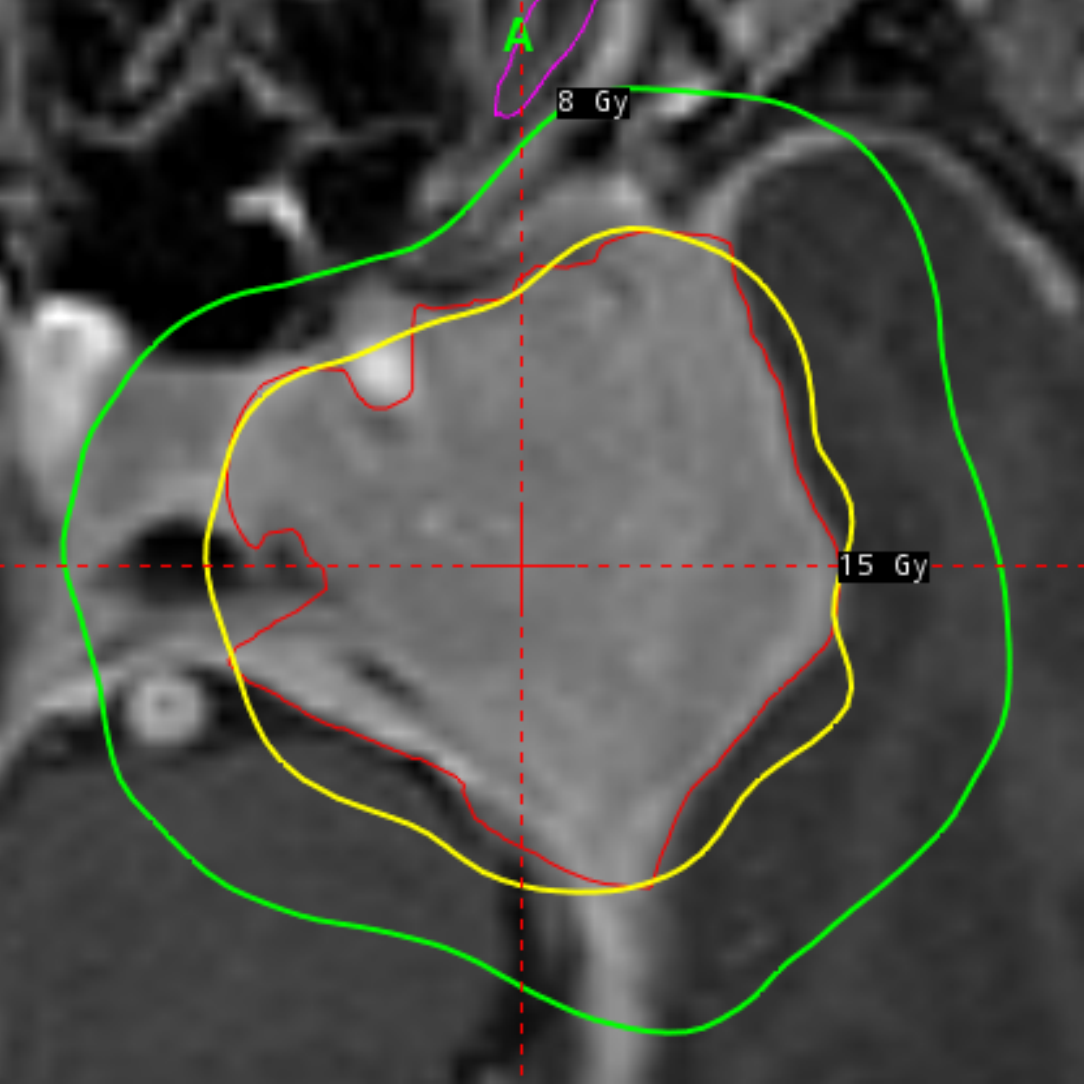}}%
\qquad
\subfigure[Promoting BOT]{%
\includegraphics[width = 0.55\textwidth]{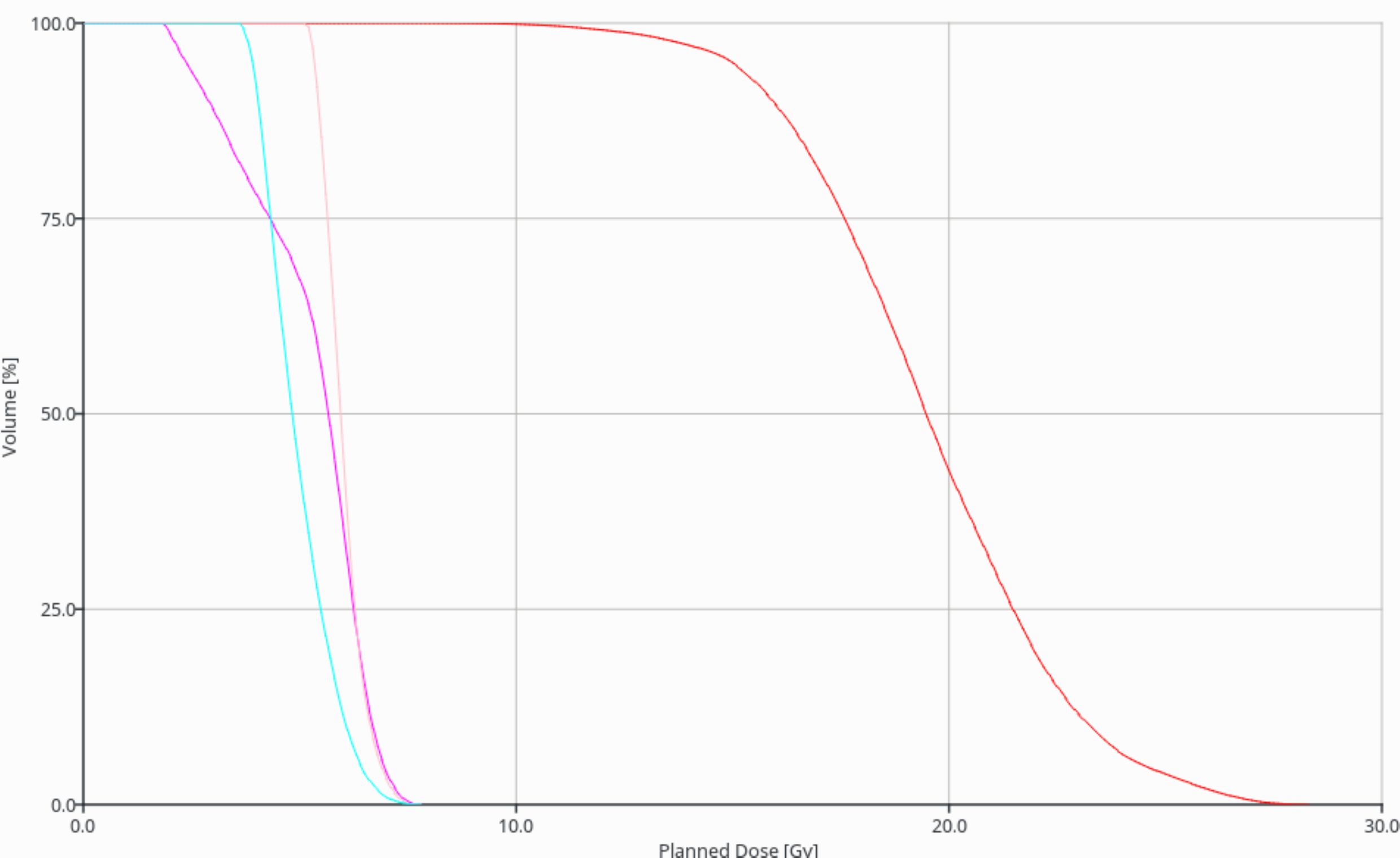}}%
\caption{Snapshots and DVHs of the reference plan, the plan promoting selectivity and the plan promoting BOT for the meningioma with adjacent OARs. The DVH for the target is in red and the three OARs in magenta, light blue and pink.}\label{fig:plans}
\end{figure}

As an example, we consider a large left-sided cavernous sinus meningioma, with a volume of  $16.2\,\textrm{cm}^3$ and with three adjacent OARs: the chiasm, the left optic nerve and the left optic tract. The prescription dose to target is 15 Gy, the dose in the inner shell encompassing the target is penalized if the dose exceeds 15 Gy. In the outer shell we penalize doses exceeding 7.5 Gy. The maximal allowed dose to all three OARs is 8 Gy, which is enforced by hard constraints. In Tab.~\ref{tab:GFRAME41}, we present two optimized plans, one promoting selectivity and one promoting short BOT, together with the clinical plan for comparison. Figure~\ref{fig:plans} shows snapshots and dose-volume histograms from Leksell GammaPlan\textsuperscript{\textregistered} for the three different plans. In the plan where short BOT is promoted, the BOT is almost halved at the expense of a moderate decrease in selectivity compared to the second plan. However, we consider both plans clinically acceptable. 
\begin{center}
\renewcommand\tabcolsep{6pt}
\begin{table}[ht!]
    \caption{Plan metrics for the reference plan and two optimized plans}\label{tab:GFRAME41}
    \begin{tabular}{ l  c  c  c  }
    \hline\hline
     & Reference plan & Promoting selectivity & Promoting BOT \\ \hline
    Coverage& 0.95 & 0.95 & 0.96  \\ 
    Selectivity & 0.88 & \textbf{0.91} & 0.78 \\ 
    Gradient index & 2.73 & 2.79 & 2.71 \\ 
    BOT (min) at 3 Gy/min & 89 & 135 & \textbf{41} \\ 
    Max dose (Lt Optic nerve) &8.95& 8.0 & 7.2 \\ 
    Max dose (Lt Optic tract) & 7.3 & 8.0 & 7.2  \\ 
    Max dose (Chiasm) & 7.3& 8.0 & 7.2\\ 
    Planning isodose & 46 & 44 & 53  \\ 
    \hline\hline
    \end{tabular}
    \end{table}
\end{center}

\subsubsection{Dose painting and homogeneous plans}
The flexibility of the proposed inverse planner opens up new possibilities for treatment planning. Here, we will exemplify both dose painting and how to create plans with a homogeneous dose distribution in the target.

Dose painting is achievable since both weights and prescription doses can be modified on a voxel-by-voxel basis. To illustrate this, we introduce a hotspot, with a volume of $1.2\, \textrm{cm}^3$, in the center of the target. The dose prescribed to the hotspot is twice the dose prescribed to the rest of the structure, i.e.~30 Gy. In the present example, we get a hotspot coverage of 99~\%.

\begin{figure}%
\centering
\subfigure[Homogeneous plan]{%
\includegraphics[width = 0.7\textwidth]{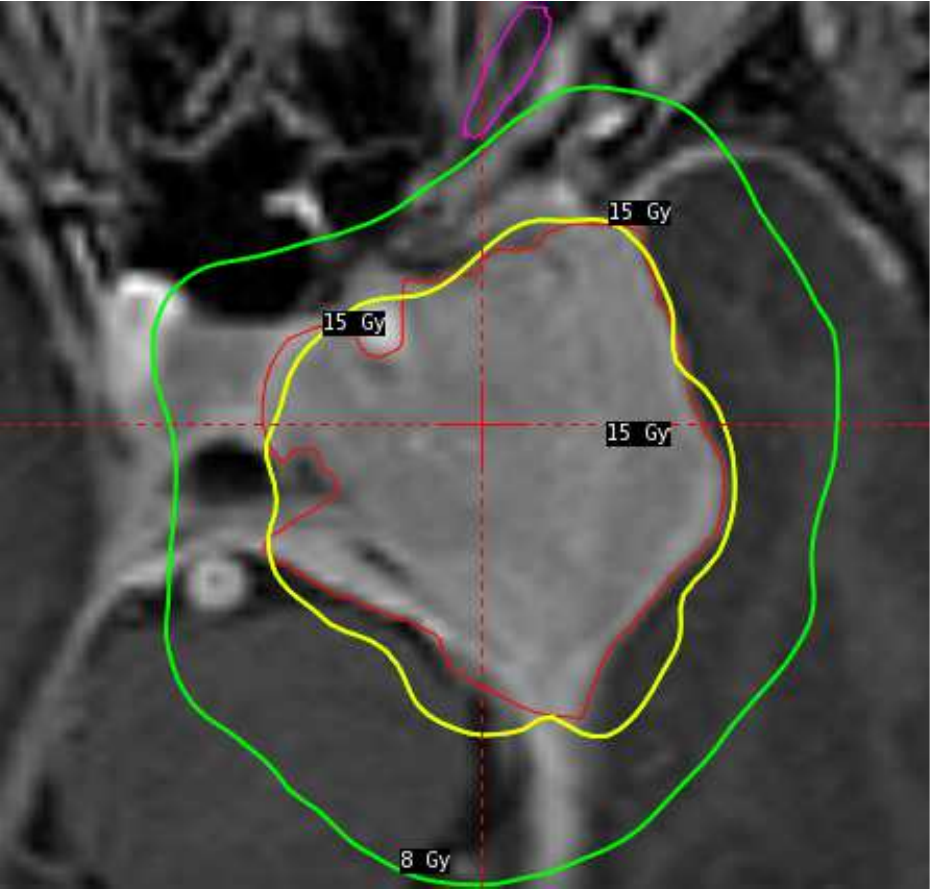}}%
\qquad
\subfigure[Plan with hotspot]{%
\includegraphics[width = 0.7\textwidth]{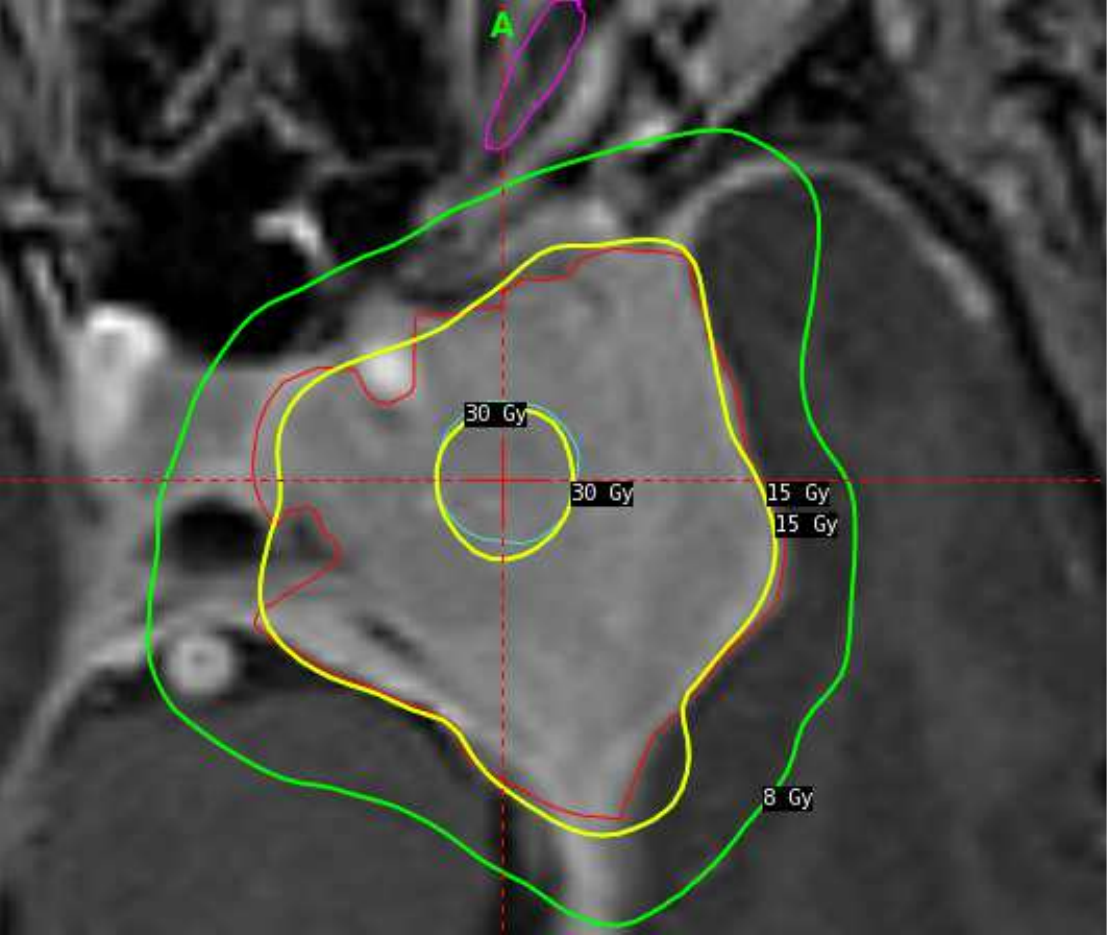}}%
\caption{Snapshots from one plan with homogeneous dose distribution and one with a hotspot}\label{fig:specialplans}
\end{figure}

It turns out that the iBOT term will in many cases favor fairly homogeneous plans. However, such plans can be further promoted by penalizing overdosage of the target. We thus penalize doses exceeding $15/0.85$ Gy to get a planning isodose of at least 85\%, which is very difficult using forward planning. In Tab.~\ref{tab:hotspot}, we present an example of a plan for the hotspot case and one plan with a homogeneous dose distribution. Overall, we consider both plans acceptable, although naturally the additional requirements result in some form of trade-off. The hotspot case has similar plan quality but longer BOT than the clinically acceptable reference plan presented in Tab.~\ref{tab:GFRAME41}. The homogeneous plan, on the other hand, trades off selectivity and we need to loosen the hard constraint on dose to OARs to achieve homogeneity. In Fig.~\ref{fig:specialplans}, we present snapshots of one plan with a homogeneous dose distribution and one plan with a hotspot. 
\begin{center}
\begin{table}[ht!]
\renewcommand\tabcolsep{6pt}
\caption{One plan with a hotspot and one homogeneous plan}\label{tab:hotspot}
    \begin{tabular}{l  c  c  }
    \hline\hline
     & Hotspot\: & Homogeneous plan \\ \hline
    Coverage& 0.97& 0.95  \\ 
    Selectivity & 0.85&  0.87 \\ 
    Gradient index & 2.9& 2.8  \\ 
    BOT (min) at 3 Gy/min & 136 & 152\\ 
    Max dose target (Gy) &  38.5& 17.7 \\ 
    Max dose Lt Optic nerve (Gy) & 8.0 & 8.0 \\ 
    Max dose Lt Optic tract (Gy) & 8.0 & 8.0  \\ 
    Max dose Chiasm (Gy) & 8.0 & 8.0\\ 
    Planning isodose &  40&  85 \\ 
    \hline\hline
    \end{tabular}
    \end{table}
\end{center}

\section{Discussion}
In this work, we have described a linear programming approach to Gamma Knife radiosurgery based on the division of the planning into three distinct phases: isocenter placement, optimization and sequencing. Our main focus has, however, been the optimization phase. We have shown that even without changing the isocenter positions, our optimization approach can find plans that dominate manual forward plans in almost 60\% of the cases investigated. Reusing the isocenter positions in this way clarifies the improvement that results directly from the optimization, but it is---of course---not how our approach would eventually be deployed. A natural next step is thus to investigate methods for isocenter placement with an emphasis on the interplay with subsequent phases.

Our optimization problem, with a cost function that is a weighted sum of multiple competing objectives, is fundamentally a multi-criteria optimization problem. A useful concept in multi-criteria optimization is that of the Pareto surface. Simply put, the Pareto surface is the set of solutions in which you cannot improve one objective without impairing another. This means one should never be satisfied with a solution that is not on the Pareto surface (Pareto optimal) unless other considerations than explicitly expressed in the cost function is taken into account. For example, our cost function only models gradient index and selectivity indirectly, which means that we cannot guarantee Pareto optimality with respect to these metrics. On the other hand, our results show that manual forward plans are, in general, not Pareto optimal either and that the proposed inverse planner can often find plans that dominate them.

For convex multi-criteria optimization problems, such as ours, every choice of weights corresponds to a Pareto optimal solution. By specifying the weights before optimization, we have treated this problem as a single-criteria optimization problem~\citep{Ehrgott05multicriteriaoptimization}. However, every Pareto optimal point corresponds to the solution of the optimization problem for a set weights on the unit simplex, i.e. such that $w\geq 0$ and  $\sum_{i=1}^n w_i=1$. In other words, the formulation as a weighted sum does not restrict the set of optimal plans that we can obtain. In practice, our normalization is not ideal since the number of active points in each structure depends on a number of factors, including the volume and the sampling strategy. Consequently the exact weight settings are unlikely to generalize from one case to another even though they may be quite similar. We thus expect that the planner will have to adjust the weights a few times before arriving at a satisfactory solution.

Although not explored here, we anticipate that further computational gains are attainable by cleverly manipulating the dose rate matrix $\Phi(r)$ and adapting the matrix algebra accordingly\cite{Cho2001,Thieke2002,Zakarian2004}. We also expect that tuning or replacing the solver, or both, could shorten the optimization times further\cite{Olafsson2006}.

\section{Conclusions}

We present the first sector-duration optimization that uses linear programming. It uses a beam-on time (BOT) penalization tailored to the Gamma Knife, which reduces the BOT by a factor of 2--3 compared to the na\"{i}ve alternative. In addition to using linear programming, we describe two techniques that reduce the problem size and thus further reduce the solution time: dualization and representative subsampling. Dualization leads to an equivalent problem that can be solved 5--20 times faster than the primal one. With representative subsampling we refer to a stochastic sampling of positions, both in the interior and on the surface of relevant structures, to use in the optimization. We show that using 10\% of the original number of voxels is enough to generate plans for which the statistical fluctuations in coverage and selectivity are below 1\% while resulting in time-savings comparable to dualization. 

The importance of different objectives, such as coverage, selectivity or beam-on time (BOT), are controlled by adjusting the weights of the corresponding terms in the cost function. In a comparison with 75 clinical plans we show that in 44 of these we can find plans that simultaneously have better selectivity, BOT and gradient index (with coverage close to 100\% in all cases) than the forward plans.     

Treatment planning for Gamma Knife has always been highly interactive. Thanks to the combination of techniques to reduce the computational cost that we have presented, it becomes possible to fit sector-duration optimization---with all its benefits---into the clinical workflow. This is our main contribution.

\section*{Disclosure of Conflicts of Interest}
All authors are employed at Elekta Instrument AB, which holds several patents related to this area. The work presented may become part of a future commercial product. 

\begin{acknowledgments}
The authors would like to thank Björn Somell and the anonymous reviewers for helpful comments and suggestions in the preparation of the manuscript. The Sharknado team---nuff said. The research was supported by the VINNOVA/ITEA3 project BENEFIT (grant 2014-00593).
\end{acknowledgments}

\appendix
\section{Explicit formulation of the primal problem}\label{appendix:explicit_primal}
Any linear programming problem can be written in the form
\begin{equation}
\begin{aligned}
& \underset{\vecx}{\text{minimize}} 
& & \vecw^t\vecx \\
& \text{subject to}& & A\vecx = b\\
&&& \vecx\geq 0,\label{eq:standardLP}
\end{aligned}
\end{equation}
where $x\in\mathbb{R}^n$ are the optimization variables and $w$, $b$ and $A\in\mathbb{R}^{p \times n}$ define the objective and constraint functions of the problem.

We introduce auxiliary variables to rewrite the hinge and iBOT functions using linear programming:
if $y^+, y^-\geq 0$ and $D - \hat{D} = (y^+ - y^-)$, then $y^+\geq (D-\hat{D})_+$ and $y^-\geq (\hat{D}-D)_+$; if $\tau_i\geq \sum_{c=1}^3 t_{isc} $ for $s=1,\ldots,8$, then $\tau_i\geq \underset{s}{\max}\sum_{c=1}^3 t_{isc}$. We may then express the primal optimization problem \eqref{eq:costfcn} in standard form by making the following identifications: 
\begin{equation}
\begin{aligned}
A &= \begin{pmatrix}
 \Phi_{\rm T} &-I&I&0&0&0&0&0&0&0\\
 \Phi_{\rm S} &0&0&-I&I&0&0&0&0&0\\
 \Phi_{\rm G} &0&0&0&0&-I&I&0&0&0\\
\Phi_{\rm O} &0&0&0&0&0&0&I&0&0\\
C &0&0&0&0&0&0&0&I&-T\\
  \end{pmatrix}, \\
x^t &=\left( \tOpt, y_\text{T}^+,y_\text{T}^-,y_\text{S}^+,y_\text{S}^-,y_\text{G}^+,y_\text{G}^-,p,q,\tau\right),\\
\vecw^t &=\left(0, 0, \frac{w_{\rm T}}{N_{\rm T} D_{\rm T}}, \frac{w_{\rm S}}{N_{\rm S} D_{\rm S}}, 0, \frac{w_{\rm G}}{N_{\rm G} D_{\rm G}}, 0, 0,0, \frac{w_{\rm BOT}}{D_\textrm{T}/\varphi_\text{cal}}\right),\\
b^t &= \left(D_{\rm T},D_{\rm S},D_{\rm G},D_{\rm O},0\right),
\end{aligned}
\label{eq:explicitPrimal}
\end{equation}
where 
\begin{align}
C&=\begin{pmatrix}
1&1&1&0&0&0&\ldots&0&0&0\\
0&0&0&1&1&1&\ldots&0&0&0\\
\vdots&\vdots&\vdots&\vdots&\vdots&\vdots&\ddots&\vdots&\vdots&\vdots\\
0&0&0&0&0&0&\ldots&1&1&1
\end{pmatrix}
= I_{N_\text{iso}}\otimes I_8\otimes (1, 1, 1)\in\mathbb{R}^{8N_\text{iso}\times 24N_\text{iso}},\\
T &=I_{N_\text{iso}}\otimes 1_{8\times 1}\in \mathbb{R}^{8N_\text{iso}\times N_\text{iso}},
\end{align}
and $\otimes$ denotes the Kronecker product.

\section{Explicit formulation of the dual problem}\label{appendix:explicit_dual}
Most modern linear programming solvers begin with a presolve step that intends to reduce the problem size \cite{Andersen1995}. However, for completeness and because we haven't encountered a solver that automatically detects the benefit of dualizing our problem, we will here carry out the dualization explicitly.

The dual problem corresponding to a linear programming problem on standard form, equation \eqref{eq:standardLP}, is\cite{Bertsimas1997,Boyd2004,Nocedal2006}
\begin{equation}
\begin{aligned}
& \underset{\nu}{\text{minimize}} 
& & b^t\nu \\
& \text{subject to}& & A^t\nu+\vecw \geq 0,\label{eq:dualstandardLP}\\
\end{aligned}
\end{equation}
where the dual variable $\nu$ is the Lagrange multiplier for the linear constraints of the primal problem. By eliminating redundant variables and grouping constraints, our problem can be stated as
\begin{equation}
\begin{aligned}
& \underset{\tilde{\nu}}{\text{minimize}} 
& & \rho^t\tilde{\nu} \\
& \text{subject to}& & B^t\tilde{\nu}\leq \sigma,\\
&& & \ell \leq \tilde{\nu} \leq \mu \,,\label{eq:ourDual}
\end{aligned}
\end{equation}
where we have introduced the following rescaled entities
\begin{equation}
\begin{aligned}
\tilde{\nu} &= (\frac{\nu_{\rm T}}{D_{\rm T}}, \frac{\nu_{\rm S}}{D_{\rm S}}, \frac{\nu_{\rm G}}{D_{\rm G}}, \frac{\nu_{\rm O}}{D_{\rm O}}, \frac{\nu_\text{iso}}{D_\textrm{T}/\varphi_\text{cal}})\,,\\
\varphi_{\rm T}&= \frac{\phi_{\rm T}}{D_{\rm T}},\,\varphi_{\rm S}= \frac{\phi_{\rm S}}{D_{\rm S}},\,\varphi_{\rm G}= \frac{\phi_{\rm G}}{D_{\rm G}},\,
\varphi_{\rm O}= \frac{\phi_{\rm O}}{D_{\rm O}},\, C_s = \frac{C}{D_\textrm{T}/\varphi_\text{cal}}.\\
\end{aligned}
\end{equation}
We may then express the dual optimization problem corresponding to equation \eqref{eq:explicitPrimal} by making the following identifications
\begin{equation}
\begin{aligned}
\rho^t &= \left(-1, 1, 1, 1, 0\right)\,,\\
B^t &= \begin{pmatrix}
    \varphi^t_{\rm T} & -\varphi^t_{\rm S} & -\varphi^t_{\rm G} & -\varphi^t_{\rm O} & -C_s^t  \\
    0 & 0 & 0 & 0 & T^t
  \end{pmatrix}\,,\\
\sigma^t &= (0,w_{\rm BOT})\,, \\
\ell^{\phantom t} &=  \left(0, 0, 0, 0, 0\right)\,,\\
\mu^{\phantom t} &= \left(\frac{w_{\rm T}}{N_{\rm T}},\frac{w_{\rm S}}{N_{\rm S}},\frac{w_{\rm G}}{N_{\rm G}},\infty,\infty \right)\,.
\end{aligned}
\end{equation}
The number of non-zero elements in $B$ is $\left(24(N_\text{T}+N_\text{S}+N_\text{G}+N_\text{O}) + 32\right)N_\text{iso}$.

\end{document}